\documentclass[preprint,12pt,authoryear]{elsarticle}

\usepackage[utf8]{inputenc}
\usepackage[T1]{fontenc}
\usepackage{amsmath,amssymb}
\usepackage{graphicx}
\usepackage{booktabs}
\usepackage{lineno}
\usepackage{hyperref}
\usepackage{subcaption}
\usepackage{float}

\journal{Journal of Monetary Economics}

\begin{document}

\begin{frontmatter}

\title{United in Currency, Divided in Growth: Dynamic Effects of Euro Adoption\tnoteref{t1}}
\tnotetext[t1]{This paper extends ``The European Union and Economic Growth: The Average Treatment Effect of Adopting the Euro,'' a chapter from the author's doctoral dissertation that applied propensity score matching and synthetic control methods. Earlier versions were presented at several conferences, and I thank participants for valuable feedback. I am grateful to Birol Kanik for helpful discussions and comments on earlier drafts. Recent advances in causal machine learning enabled me to revisit the research question and estimate the dynamic heterogeneous treatment effects that were not feasible with earlier methods.}

\author[aws]{Harry Aytug}
\ead{haytug@amazon.com}

\address[aws]{Amazon Web Services}

\begin{abstract}
Does euro adoption affect long-run economic growth? Existing evidence is mixed, reflecting limited treated countries, long horizons that challenge inference, and heterogeneity across member states. We estimate causal dynamic and heterogeneous treatment effects using Causal Forests with Fixed Effects (CFFE), a machine-learning approach that combines causal forests with two-way fixed effects. Under a conditional parallel-trends assumption, we find that euro adoption reduced annual GDP growth by 0.3-0.4 percentage points on average. Effects emerge shortly after adoption and stabilize after roughly a decade.

Average effects mask substantial heterogeneity. Countries with lower initial GDP per capita experience larger and more persistent growth shortfalls than core economies. Weaker consumption and productivity growth contribute to the overall effect, while improvements in net exports partially offset these declines.

A two-country New Keynesian DSGE model with hysteresis generates qualitatively similar patterns: one-size-fits-all monetary policy and scarring mechanisms produce larger output losses under monetary union than under flexible exchange rates. By jointly estimating dynamic and heterogeneous treatment effects, the analysis highlights the importance of country characteristics in assessing the long-run consequences of monetary union.
\end{abstract}


\end{frontmatter}

%\linenumbers  % Uncomment for line numbers during review

%==============================================================================
% INTRODUCTION
%==============================================================================
\section{Introduction}

Whether the adoption of a common currency affects long-run economic growth remains a central and unresolved question in international and macroeconomic economics. Standard theory suggests that monetary arrangements should not influence long-run real outcomes, as nominal exchange rates and monetary policy are neutral in the steady state. At the same time, a large body of work emphasizes that adjustment dynamics matter in practice: in the presence of nominal rigidities, asymmetric shocks, and limited factor mobility, the loss of monetary autonomy may shape economic performance over extended periods. Despite extensive research, existing studies struggle to identify long-run growth effects in settings with few treated countries and substantial cross-country heterogeneity.

The European Monetary Union provides a unique setting in which to study these issues. Since 1999, twenty European countries have adopted the euro, while other otherwise similar economies---both within and outside the European Union---have retained national currencies. Euro adoption represents a permanent shift in the policy environment, involving the loss of independent monetary and exchange rate policy, fiscal constraints under common rules, and exposure to a common monetary stance set at the union level. These features make the euro a natural laboratory for studying how monetary integration interacts with growth over the medium and long run.

Despite its importance, empirical evidence on the growth effects of euro adoption remains mixed. Early studies were constrained by limited post-adoption data, while later analyses were complicated by the global financial crisis and the European sovereign debt crisis, which affected member states unevenly. Synthetic control studies document substantial cross-country variation, but typically focus on individual countries and summarize effects over long post-treatment windows. Event-study and difference-in-differences approaches offer a dynamic perspective, but often impose restrictive functional forms and suffer from imprecise inference at long horizons, particularly in settings with a small number of treated units. These divergent findings largely reflect differences in identification strategies, time horizons, and the ability to account for heterogeneous treatment effects.

This paper contributes to the literature by estimating dynamic and heterogeneous growth effects of euro adoption within a unified panel-data framework. We employ Causal Forests with Fixed Effects (CFFE), a flexible estimator that allows treatment effects to vary with both time since adoption and pre-treatment country characteristics, while controlling for unobserved country- and time-specific heterogeneity. Rather than imposing parametric adjustment paths, the approach lets the data inform how growth responses evolve over time and across economies. Identification relies on a conditional parallel trends assumption: within each leaf of the forest, treated and control countries would have followed similar growth paths absent euro adoption.

Rather than replacing classical approaches, the CFFE framework complements existing methods by offering a flexible way to study long-horizon dynamics in settings with limited treated units. Unlike synthetic control studies that estimate a single long-run difference between treated and synthetic counterfactual units, CFFE traces the entire trajectory of effects from adoption through the present. Unlike classical two-way fixed effects event studies that impose linearity and can yield imprecise estimates at long horizons, the forest-based estimator provides built-in regularization that yields more stable inference even 20 years post-adoption. The method also accommodates treatment effect heterogeneity, revealing which country characteristics predict divergent adjustment paths.

Our analysis draws on panel data covering 20 eurozone countries and 15 control economies from 1970 to 2023. The treatment group includes the 11 founding members who adopted the euro in 1999, plus nine countries that joined subsequently: Greece (2001), Slovenia (2007), Cyprus and Malta (2008), Slovakia (2009), Estonia (2011), Latvia (2014), Lithuania (2015), and Croatia (2023). Controls comprise EU members that opted out (Denmark, Sweden, the UK until Brexit) and other OECD economies. We measure growth using real GDP per capita from the Penn World Tables, supplemented with macroeconomic indicators from the World Bank.

The results reveal patterns that help explain divergent findings in the literature. We find annual growth declines of 0.3--0.4 percentage points, with effects stabilizing after approximately ten years. This average, however, masks substantial heterogeneity. Periphery economies---Greece, Ireland, Italy, Portugal, and Spain---experienced larger growth shortfalls of approximately 0.5 percentage points, while core members saw smaller impacts around 0.3 percentage points. Initial GDP per capita alone explains around 45\% of cross-country heterogeneity, suggesting that less developed economies faced greater adjustment challenges.

Mechanism analysis reveals that both consumption and productivity growth decline following euro adoption, while net exports improve, partially offsetting the contraction in domestic demand. Investment effects were smaller and less persistent. The decline in consumption is consistent with the loss of monetary policy flexibility constraining domestic demand management, while the productivity slowdown proves difficult to disentangle from broader European trends.

These findings help clarify why existing studies reach divergent conclusions. Studies finding no average effect are not necessarily incorrect; they may report a mean that combines positive and negative experiences. The synthetic control literature's focus on individual countries captures heterogeneity but sacrifices statistical power and dynamic inference. Classical event studies impose structure that may obscure nonlinear adjustment paths. By combining the strengths of these approaches---dynamic estimation, heterogeneity analysis, and formal inference---CFFE offers a complementary perspective on how euro adoption reshaped growth trajectories across the currency union, contributing to the literature on monetary integration and optimal currency areas.

The remainder of the paper proceeds as follows. Section 2 reviews the literature on euro growth effects and positions our methodological contribution. Section 3 describes the CFFE estimator and its application to dynamic treatment effects. Section 4 presents the data and sample construction. Section 5 reports results on average effects, heterogeneity, and mechanisms. Section 6 discusses implications and limitations. Section 7 concludes.

%==============================================================================
% LITERATURE REVIEW
%==============================================================================
\section{Literature Review}

Research on exchange rate regimes and economic growth spans theoretical and empirical traditions that reach conflicting conclusions. This section reviews the broader literature on exchange rate regime effects before turning to euro-specific studies and positioning our methodological contribution.

\subsection{Theoretical Foundations}

Standard macroeconomic theory offers ambiguous predictions about exchange rate regime effects on growth. In frictionless models with complete markets and flexible prices, nominal regime choice is irrelevant for long-run real allocations---money is neutral in steady state. \citet{barro1983rules} extend this logic to policy credibility, arguing that attempts to exploit the Phillips curve through monetary expansion or currency devaluation produce higher inflation rather than sustained output gains. From this perspective, the choice between fixed and flexible regimes should be neutral for long-run growth.

Yet theoretical arguments cut both ways. \citet{friedman1953case} emphasizes that flexible exchange rates can absorb external shocks, allowing faster adjustment than the slow price-level changes required under fixed rates. In a world of Keynesian price rigidities, this adjustment advantage could translate into growth benefits. Conversely, fixed regimes may stimulate investment and trade by reducing policy uncertainty and price volatility \citep{mckinnon2004optimum}. \citet{ghosh1997does} document patterns consistent with this tradeoff: fixed regimes are associated with higher investment, while flexible regimes correlate with faster productivity growth. According to the Solow growth model, output growth derives from either factor accumulation or total factor productivity, suggesting that regime choice may affect the composition rather than the level of growth. Importantly, euro adoption differs from a generic fixed exchange rate regime: it also changes the lender of last resort, eliminates redenomination risk, deepens financial integration, and imposes fiscal constraints---an institutional bundle whose effects may exceed those of the exchange rate peg alone.

\citet{levy2003fear} provide a comprehensive framework showing how exchange rate regimes matter for growth through multiple channels, though the sign of the net effect remains theoretically ambiguous. Empirical estimates remain sensitive to regime classification, sample composition, and identification strategy.

\subsubsection{Why Might the Euro Affect Real Growth?}

The classical neutrality proposition---that nominal variables cannot affect real outcomes in the long run---requires qualification in the context of monetary union. Several channels may generate persistent real effects from euro adoption, each with testable implications for consumption, investment, productivity, and net exports that we examine in Section 5.

The first channel operates through adjustment mechanism costs. When countries face asymmetric shocks, flexible exchange rates provide a rapid adjustment mechanism. Under monetary union, adjustment must occur through internal devaluation (wage and price cuts) or factor mobility. Both alternatives are slower and more costly than exchange rate adjustment. If shocks are frequent or large, the cumulative adjustment costs can reduce trend growth through hysteresis and scarring: prolonged unemployment erodes skills, delayed investment becomes foregone investment, and discouraged workers exit the labor force permanently. This channel predicts larger effects for countries with rigid labor markets, high pre-euro inflation differentials, or low business cycle correlation with the eurozone core.

A second channel involves fiscal policy constraints. The Stability and Growth Pact imposes deficit and debt limits, but the binding constraint during crises came from market discipline rather than rules: sovereign spreads widened sharply for periphery countries, and the absence of a monetary backstop until 2012 amplified fiscal stress. When monetary policy is unavailable and market access is constrained, countries may experience deeper and longer recessions. Hysteresis effects can then translate temporary demand shortfalls into permanent supply-side damage. This channel predicts larger effects for countries with high pre-crisis debt or greater exposure to sovereign spread shocks.

Third, one-size-fits-all monetary policy may generate persistent effects. ECB policy is set for the eurozone aggregate, which may be inappropriate for individual members. Countries with higher inflation or faster growth may face interest rates that are too low, fueling credit booms, housing price appreciation, and current account deficits. Countries in recession may face rates that are too high, prolonging downturns. Persistent policy misalignment can affect long-run growth through investment distortions and sectoral reallocation toward non-tradables during boom periods.

Finally, even if money is neutral in the very long run, transition dynamics can be prolonged. Our 20-year sample may capture an extended adjustment period rather than a permanent growth reduction. The stabilization of effects after year 10 is consistent with this interpretation: the marginal growth effect attenuates even as the cumulative level gap persists. Countries may eventually adapt to monetary union, but the adjustment costs are substantial and long-lasting.

These channels provide a framework for interpreting our estimates. Monetary neutrality serves as a theoretical benchmark; frictions and institutional constraints imply deviations from that benchmark whose magnitude and persistence are ultimately empirical questions.

\subsection{Exchange Rate Regimes and Growth: Prior Evidence}

The empirical relationship between exchange rate regimes and growth has been studied extensively, with conflicting results. Some studies find fixed regimes conducive to growth \citep{mundell1995exchange, dubas2005exchange}, others favor flexible regimes \citep{levy2003fear, eichengreen2003capital}, and still others find no significant relationship \citep{baxter1989business, husain2005exchange}. These divergent findings reflect differences in regime classification schemes, sample composition across developing and advanced economies, the choice between level and growth-rate outcomes, and the treatment of endogeneity. In advanced economies, regime changes are rare and often coincide with broader institutional reforms, making identification particularly challenging.

A key finding from this literature is that regime effects may differ between developing and industrial countries. \citet{huang2004exchange} find that exchange rate regimes matter for developing countries but not for developed economies. This distinction is relevant for the eurozone, which comprises exclusively developed economies---though even within this group, convergence status may matter. Later adopters such as the Baltic states and Slovakia were still converging toward Western European income levels, and may exhibit growth dynamics more similar to emerging Europe than to the eurozone core.

The central methodological challenge is endogeneity: countries do not randomly select into fixed or flexible arrangements. Euro adoption presents specific endogeneity concerns: the decision to join reflected political economy considerations, satisfaction of convergence criteria that themselves depended on prior economic performance, and---for later adopters---bundling with EU accession. Moreover, anticipation effects complicate identification: markets, firms, and policymakers adjusted behavior well before the formal adoption date, potentially shifting some effects into the pre-treatment period. The synthetic control method and difference-in-differences designs offer identification strategies that exploit the specific timing of regime changes, and have become prominent in euro-specific studies.

\subsection{Synthetic Control Studies of the Euro}

The synthetic control method (SCM), introduced by \citet{abadie2003economic} and refined in subsequent work, constructs counterfactual outcomes as weighted combinations of control units. Applied to euro adoption, SCM estimates what GDP would have been had a country not joined the currency union. Most SCM studies target GDP per capita levels rather than growth rates; our analysis focuses on growth rates for two reasons. First, growth rates map directly to dynamic adjustment paths and allow us to distinguish between temporary transition costs and persistent growth-rate wedges. Second, growth rates connect naturally to mechanism variables (consumption growth, investment growth, productivity growth), enabling consistent decomposition of the headline effect.

A recent comprehensive study by \citet{gabriel2024euro} examines all eurozone members individually, using OECD countries as donors and GDP per capita as the outcome over a 20-year post-treatment horizon. They find substantial heterogeneity: some countries (notably Germany and the Netherlands) appear to have benefited from euro membership, while others (particularly Italy and Portugal) experienced significant growth shortfalls relative to their synthetic counterparts. The average effect across countries is close to zero, but this masks divergent experiences.

\citet{lin2017euro} apply SCM to a subset of founding members, finding mixed results that depend heavily on the choice of donor pool and pre-treatment fit---sensitivity that motivates our panel-based approach with explicit covariate adjustment. \citet{lucke2022euro} extend the analysis to later adopters, documenting that Eastern European entrants generally performed better than their synthetic controls in the years immediately following adoption. However, for these countries, euro adoption closely followed EU accession, making it difficult to separate the two treatments---a confound that also affects our analysis and that we address through robustness checks. SCM has also been applied to other European integration policies: \citet{aytug2017customs} use SCM to evaluate the EU--Turkey Customs Union, finding sizable trade effects relative to a synthetic counterfactual, which illustrates both the method's applicability to integration questions and the sensitivity of results to donor pool construction.

SCM offers transparent counterfactual construction and handles the fundamental problem of causal inference elegantly. Its limitations for our purposes are threefold. First, while SCM can trace post-treatment paths, effects are often summarized as a single long-run difference rather than dynamic adjustment trajectories with formal inference at each horizon. Second, inference relies on permutation tests that may lack power with small donor pools. Third, aggregating individual country estimates into an overall effect requires additional assumptions about weighting. Recent extensions---generalized SCM, panel synthetic control, and matrix completion methods---address some of these limitations, and our CFFE approach can be viewed as part of this broader toolkit for panel causal inference with heterogeneous effects.

\subsection{Difference-in-Differences and Event Studies}

The difference-in-differences (DiD) framework compares changes in outcomes between treated and control groups before and after treatment. Standard TWFE designs applied to euro adoption have produced mixed or imprecise estimates. \citet{ioannatos2018euro}, for example, finds no statistically significant average effect on GDP growth, though confidence intervals are wide enough to encompass economically meaningful effects in either direction.

Event study designs extend DiD by estimating separate effects at each time horizon relative to treatment. This approach has become standard in applied microeconomics following methodological advances by \citet{sun2021estimating} and \citet{callaway2021difference}. Applied to macroeconomic settings, event studies face additional challenges beyond those in micro panels: cross-sectional dependence from common shocks, small numbers of treated units, and strong serial correlation in outcomes.

First, with staggered adoption timing, two-way fixed effects estimators can produce biased estimates when treatment effects are heterogeneous across cohorts or time \citep{goodman2021difference}. The euro setting exhibits both features: the 1999 cohort differs from later adopters, and effects likely vary with economic conditions at adoption. TWFE can misweight comparisons by using already-treated countries as implicit controls, contaminating estimates. We report TWFE results for comparability but do not rely on them for our main conclusions.

Second, standard errors in event studies often explode at long horizons due to declining sample sizes and increasing variance. This problem is particularly acute for euro adoption, where we observe some countries for over 20 years post-treatment but others for much shorter periods. Our CFFE approach provides more stable inference through regularization, but regularization is not free: we validate our uncertainty estimates through block bootstrap by country, placebo adoption dates, and leave-one-out sensitivity analysis.

Third, classical event studies are flexible in event time but struggle to incorporate high-dimensional heterogeneity. Estimating interactions between event time and country characteristics (k $\times$ X) quickly exhausts degrees of freedom with limited treated units. Causal forests handle this dimensionality naturally by partitioning the covariate space adaptively.

\subsection{Broader EU Integration and Core-Periphery Asymmetries}

The euro growth debate sits within a larger literature on European integration. \citet{campos2019economic} find generally positive effects of EU membership on GDP per capita using SCM, but the incremental effect of euro adoption conditional on EU membership remains unclear. Studies of trade effects consistently find that the euro increased bilateral trade among members \citep{rose2000one, glick2016currency}, though whether trade gains translated into growth gains is less clear. This literature highlights a key identification challenge: for late adopters, EU membership and euro adoption are tightly linked, and trade integration may be EU-driven rather than euro-driven. We address this by presenting results separately for founding members (cleanest euro timing), by controlling for years since EU accession, and by conducting placebo tests assigning ``euro adoption'' to non-euro EU members around 1999 to test for common EU shocks.

A recurring theme is the distinction between core and periphery eurozone members. We operationalize this distinction using initial GDP per capita, which correlates strongly with other proposed measures including current account positions, labor market rigidity, and pre-crisis fiscal space. Core economies entered monetary union with stronger fiscal positions and more flexible labor markets, while periphery economies faced greater structural challenges. \citet{hall2012euro} argues that the euro's institutional design favored export-led growth models prevalent in core economies. The elimination of exchange rate risk also triggered large capital flows from core to periphery, financing consumption and housing booms that reversed sharply during the 2010--2012 sovereign debt crisis---a sudden stop dynamic that is central to understanding differential outcomes within a monetary union. Empirical studies confirm heterogeneous effects along core-periphery lines: \citet{gabriel2024euro} find that synthetic control estimates are more negative for periphery countries.

\subsection{Methodological Gaps}

Despite extensive research, several gaps remain. First, few studies jointly estimate dynamic responses and systematic heterogeneity using a unified framework with formal inference, especially with small treated samples. Interactive fixed effects and factor models offer one approach to policy evaluation in panels with unobserved confounders. Generalized synthetic control and matrix completion methods extend SCM to multiple treated units. Modern staggered-adoption DiD estimators handle heterogeneous effects across cohorts. Yet none of these methods is designed to discover which pre-treatment characteristics predict treatment effect heterogeneity in a data-driven way while simultaneously tracing dynamic adjustment paths.

Second, the literature lacks a unified framework for understanding why effects vary across countries. Individual case studies document divergent experiences, but systematic analysis of which pre-treatment characteristics predict adjustment paths---and how those paths evolve over time---is limited. Our goal is not merely to document that heterogeneity exists, but to identify which observable features predict the trajectory of effects, reducing concerns about ``bad controls'' by focusing exclusively on pre-determined variables.

Third, mechanism analysis remains underdeveloped. Studies document effects on GDP but rarely decompose these into consumption, investment, and trade channels with the same rigor applied to the headline growth estimates. Moreover, monetary-union-specific mechanisms---credit growth, real effective exchange rates, unit labor costs, and current account dynamics---are rarely examined in a unified causal framework.

Our CFFE approach addresses these gaps by estimating dynamic treatment effects that vary with country characteristics, applying the same framework to mechanism variables including productivity and financial channels. The forest-based estimator provides more stable inference at long horizons than classical event studies, but regularization is not a free lunch: we complement forest-based inference with placebo tests, block bootstrap, and comparisons to alternative estimators to validate our uncertainty quantification.

To strengthen identification and connect our heterogeneity findings to optimal currency area theory, we pursue several additional analyses. First, we benchmark CFFE against modern staggered-adoption DiD estimators and interactive fixed effects models. Second, we conduct leave-one-country-out sensitivity analysis, particularly important given the small number of founding members. Third, we assign placebo ``euro adoption'' to non-euro EU members (Denmark, Sweden, UK) around 1999 to test for spurious effects from common EU shocks. Fourth, we extend heterogeneity analysis beyond initial GDP per capita to include pre-euro inflation differentials, fiscal space, and trade integration with the eurozone---features that map directly to OCA theory predictions. The next section describes the methodology in detail.

%==============================================================================
% METHODOLOGY
%==============================================================================
\section{Methodology}

This section describes the Causal Forests with Fixed Effects (CFFE) estimator and its application to dynamic treatment effect estimation. We begin by defining the causal estimand, then describe the causal forest framework, explain how fixed effects are incorporated, and show how event time enters as a feature to recover dynamic effects.

\subsection{Setup, Notation, and Estimand}

Consider a panel of $N$ countries observed over $T$ time periods. Let $Y_{it}$ denote the outcome---annual GDP per capita growth rate---for country $i$ in year $t$. Define the treatment indicator $D_{it} = 1$ if country $i$ has adopted the euro by year $t$, and $D_{it} = 0$ otherwise. Let $T_i^{\text{euro}}$ denote the year country $i$ adopted the euro, with $T_i^{\text{euro}} = \infty$ for never-treated controls.

Event time is defined as:
\begin{equation}
k_{it} = t - T_i^{\text{euro}}
\end{equation}
for treated countries, representing years since adoption. For never-treated controls, $k_{it}$ is undefined (or set to a placeholder value).

Our target estimand is the dynamic average treatment effect on the treated (ATT) at event time $k$:
\begin{equation}
\text{ATT}(k) = \mathbb{E}[Y_{it}(1) - Y_{it}(0) \mid D_i = 1, k_{it} = k]
\end{equation}
where $Y_{it}(1)$ and $Y_{it}(0)$ denote potential growth rates under euro adoption and the counterfactual of remaining outside the eurozone. This estimand captures the causal effect of euro adoption on the annual growth rate, $k$ years after adoption, for countries that adopted the euro. Cumulating $\text{ATT}(k)$ over horizons yields the implied effect on GDP levels.

Because euro adoption was widely anticipated---markets, firms, and policymakers adjusted interest rates, fiscal policy, and capital flows well before 1999---our identifying assumption is not strict no-anticipation. Rather, we assume that conditional on country and year fixed effects and pre-treatment characteristics, remaining deviations in growth after adoption reflect causal effects of operating within the monetary union. Pre-treatment coefficients in our event study may therefore reflect anticipation rather than pure placebo, and we interpret them accordingly.

Let $X_i$ denote a vector of pre-treatment country characteristics measured before euro adoption. These include initial GDP per capita, trade openness, investment share, and human capital. The feature vector for estimation combines event time and country characteristics: $\tilde{X}_{it} = (k_{it}, X_i)$.

Country fixed effects $\alpha_i$ absorb time-invariant factors including institutions, geography, and culture. Year fixed effects $\gamma_t$ absorb common shocks such as global recessions, oil price movements, and the 2008 financial crisis. However, time fixed effects remove common shocks but not heterogeneous exposure to global shocks---countries may respond differently to the same global event. We address this limitation through robustness checks including interactive fixed effects and leave-one-out analysis.

\subsection{Causal Forests}

Causal forests, introduced by \citet{wager2018estimation}, extend random forests to estimate heterogeneous treatment effects. The method partitions the covariate space into regions where treatment effects are approximately constant, then estimates local average treatment effects within each region. Crucially, identification still relies on selection-on-observables: causal forests assume unconfoundedness conditional on the included covariates. The method provides flexible nonparametric estimation of heterogeneous effects, but does not create exogenous variation where none exists.

The target estimand is the conditional average treatment effect (CATE):
\begin{equation}
\tau(x) = \mathbb{E}[Y_i(1) - Y_i(0) \mid X_i = x]
\end{equation}
where $Y_i(1)$ and $Y_i(0)$ denote potential outcomes under treatment and control.

Causal forests estimate $\tau(x)$ by:
\begin{enumerate}
\item Growing an ensemble of trees, each trained on a bootstrap sample
\item At each split, choosing the variable and threshold that maximizes heterogeneity in treatment effects across child nodes
\item Using ``honest'' estimation: one subsample determines tree structure, another estimates leaf effects
\item Aggregating predictions across trees to obtain $\hat{\tau}(x)$
\end{enumerate}

The honest splitting procedure ensures valid inference by separating the data used for partitioning from the data used for estimation within partitions.

\subsection{Incorporating Fixed Effects}

Standard causal forests assume unconfoundedness conditional on observed covariates. In panel settings, unobserved country-specific factors (institutions, geography, culture) and time-specific shocks (global recessions, oil prices) may confound the treatment-outcome relationship. Two-way fixed effects address this concern in linear models; we extend the approach to causal forests following \citet{kattenberg2023causal}.

CFFE residualizes outcomes and treatments on fixed effects within each tree node. Specifically, for observations falling in node $\ell$, we compute:
\begin{align}
\tilde{Y}_{it}^{(\ell)} &= Y_{it} - \hat{\alpha}_i^{(\ell)} - \hat{\gamma}_t^{(\ell)} \\
\tilde{D}_{it}^{(\ell)} &= D_{it} - \hat{\pi}_i^{(\ell)} - \hat{\rho}_t^{(\ell)}
\end{align}
where $\hat{\alpha}_i^{(\ell)}$ and $\hat{\gamma}_t^{(\ell)}$ are country and year fixed effects estimated using only observations in node $\ell$, and similarly for the treatment residuals.

Node-level residualization is crucial. Global residualization (computing fixed effects once using all data) would impose that the relationship between fixed effects and outcomes is constant across the covariate space. Node-level residualization allows this relationship to vary, accommodating settings where, for example, country fixed effects matter more for some types of countries than others.

The identifying assumption becomes conditional parallel trends: within each region of the covariate space defined by the forest, treated and control units would have followed parallel outcome paths absent treatment. This is weaker than requiring parallel trends globally, but it remains an assumption. We do not claim that CFFE solves endogeneity---euro adoption is highly non-random, correlated with convergence criteria, EU accession timing, political commitment, and growth expectations. Rather, given the constraints of macro data with few treated units and no natural experiment, CFFE allows flexible modeling of heterogeneous dynamic responses under a conditional parallel trends assumption. We complement the main estimates with extensive placebo tests, leave-one-out analysis, and comparisons to alternative estimators to assess robustness.

\subsection{Dynamic Effects via Event Time Features}

The key innovation for dynamic effect estimation is including event time $k_{it}$ as a feature in the causal forest. This allows the forest to learn how treatment effects vary with time since adoption.

The model can be written as:
\begin{equation}
Y_{it} = \alpha_i + \gamma_t + \tau(k_{it}, X_i) \cdot D_{it} + \varepsilon_{it}
\end{equation}
where $\tau(k, X)$ is the treatment effect for a country with characteristics $X$ at event time $k$. The causal forest learns this function nonparametrically.

To recover the average dynamic effect at horizon $k$, we average predictions over all treated observations at that event time:
\begin{equation}
\hat{\tau}(k) = \frac{1}{N_k} \sum_{i: k_{it} = k} \hat{\tau}(k, X_i)
\end{equation}
where $N_k$ is the number of country-year observations at event time $k$.

Because growth is persistent and cumulates into levels, estimated effects at different horizons are not independent causal objects. The $\hat{\tau}(k)$ should be interpreted as deviations from counterfactual growth paths at each horizon, not as independent causal shocks. A persistent negative effect on growth rates implies a widening gap in GDP levels over time.

This approach offers several advantages over classical event studies. The forest learns the shape of $\tau(k)$ from data rather than imposing linearity or other parametric restrictions, providing flexibility that parametric approaches lack. Random forest averaging provides implicit regularization, yielding stable estimates even at long horizons where classical event studies suffer from large standard errors. Finally, the dependence of $\tau$ on $X$ is learned jointly with the dependence on $k$, revealing which country characteristics predict divergent adjustment paths.

\subsection{Inference}

Confidence intervals for $\hat{\tau}(k)$ are constructed using the forest-based variance estimator of \citet{wager2018estimation}. The estimator exploits the fact that honest forests produce asymptotically normal predictions with variance that can be estimated from the forest structure.

However, the asymptotic theory underlying these confidence intervals assumes approximately i.i.d. observations and large sample sizes. Our setting violates these assumptions: we have only 20--35 countries with strong serial correlation within countries, cross-sectional dependence from common shocks, and one major global event (the 2008 financial crisis) that affected all units. Standard forest-based inference may therefore be unreliable.

We address this concern through multiple approaches. First, for cluster-robust inference at the country level, we modify the variance estimator to account for within-country correlation. Let $\hat{\tau}_{it}$ denote the predicted treatment effect for observation $(i,t)$. The cluster-robust variance of $\hat{\tau}(k)$ is:
\begin{equation}
\widehat{\text{Var}}[\hat{\tau}(k)] = \frac{1}{N_k^2} \sum_{i} \left( \sum_{t: k_{it} = k} (\hat{\tau}_{it} - \hat{\tau}(k)) \right)^2
\end{equation}
which sums squared deviations within countries before aggregating across countries.

Second, we implement a block bootstrap at the country level: we resample entire country time series (not individual observations) and re-estimate the model on each bootstrap sample. This preserves the within-country dependence structure and provides a more conservative assessment of uncertainty.

Third, we conduct extensive robustness exercises reported in Section 6: leave-one-country-out analysis to assess sensitivity to individual countries, placebo adoption dates assigned to non-euro EU members, and comparisons to alternative estimators. CFFE estimates that fall within the confidence envelopes of these alternative approaches provide reassurance that our findings are not statistical artifacts.

\subsection{Identification Assumptions}

We summarize the key identifying assumptions underlying our analysis:

\begin{quote}
\textit{Conditional Parallel Trends}: Conditional on country fixed effects, year fixed effects, and pre-treatment covariates, euro-adopting and non-adopting countries would have followed parallel growth paths in the absence of euro adoption.
\end{quote}

This assumption is fundamentally untestable, but we provide indirect evidence through pre-treatment placebo coefficients and robustness to alternative specifications. The assumption does not require that euro adoption was randomly assigned---it clearly was not, as adoption reflected political decisions, satisfaction of convergence criteria, and EU membership status. Rather, we assume that after conditioning on observables and fixed effects, remaining variation in adoption timing is uncorrelated with future growth innovations.

We acknowledge that this assumption may be violated if, for example, countries adopted the euro precisely when they expected future growth to diverge from controls for reasons unrelated to the euro itself. The extensive robustness analysis in Section 6 is designed to probe the sensitivity of our conclusions to such violations.

\subsection{Comparison Estimators}

To assess robustness and benchmark CFFE against established methods, we estimate several alternative specifications.

The classical TWFE event study estimates:
\begin{equation}
Y_{it} = \alpha_i + \gamma_t + \sum_{k=-10}^{20} \beta_k \cdot \mathbf{1}[k_{it} = k] + \varepsilon_{it}
\end{equation}
with $k = -1$ as the reference period and standard errors clustered at the country level. This specification is transparent but may suffer from bias under heterogeneous treatment effects with staggered adoption.

The interaction-weighted estimator of \citet{sun2021estimating} addresses heterogeneity bias in staggered DiD by estimating cohort-specific effects and aggregating with appropriate weights. We implement this using the 1999 founders and later adopters as separate cohorts.

The doubly-robust estimator of \citet{callaway2021difference} combines outcome regression and propensity score weighting, providing consistent estimates under either correct specification. We report group-time average treatment effects and their aggregation to dynamic effects.

Following \citet{bai2009panel}, we also estimate an interactive fixed effects model that allows for heterogeneous exposure to common shocks:
\begin{equation}
Y_{it} = \alpha_i + \lambda_i' f_t + \tau D_{it} + \varepsilon_{it}
\end{equation}
where $\lambda_i$ are country-specific factor loadings and $f_t$ are common factors. This addresses the concern that time fixed effects remove common shocks but not differential exposure to those shocks.

We compare CFFE estimates to these alternatives in Section 5. Agreement across methods strengthens confidence in our findings; disagreement motivates investigation of the sources of divergence.

\subsection{Implementation Details}

We implement CFFE using the \texttt{causalfe} Python package \citep{aytug2026causalfe}.\footnote{\citet{kattenberg2023causal} provide an R implementation.} Key hyperparameters include:
\begin{itemize}
\item Number of trees: 500
\item Minimum leaf size: 30 observations
\item Honesty: enabled (separate subsamples for splitting and estimation)
\item Fixed effects: node-level residualization on country and year
\end{itemize}

\subsection{Heterogeneity Features}

The feature matrix includes event time $k$ and pre-treatment country characteristics. All heterogeneity features are measured prior to euro adoption, ensuring they are not themselves affected by treatment---a crucial requirement for valid causal inference that avoids ``bad control'' bias.

We select features that proxy for specific theoretical mechanisms from the optimal currency area literature:

\begin{center}
\begin{tabular}{ll}
\toprule
\textbf{Feature} & \textbf{Theoretical Mechanism} \\
\midrule
Initial GDP per capita & Convergence dynamics, structural readiness \\
Trade openness & Integration gains, exposure to external shocks \\
Investment share & Capital accumulation capacity \\
Human capital index & Adjustment capacity, labor market flexibility \\
\bottomrule
\end{tabular}
\end{center}

Initial GDP per capita captures convergence dynamics: less developed economies may face greater adjustment challenges but also have more room for catch-up growth. Trade openness proxies for integration gains from reduced transaction costs but also exposure to asymmetric trade shocks. Investment share reflects capital accumulation patterns that may interact with interest rate convergence. Human capital captures adjustment capacity through labor market flexibility and skill-based reallocation.

All continuous features are standardized before estimation. Given the small number of treated units---approximately 11 founding members provide the cleanest identification, with later adopters confounded by EU accession---heterogeneity estimates should be interpreted as descriptive patterns suggesting which country characteristics correlate with divergent adjustment paths, rather than as precisely estimated structural causal parameters. We assess robustness through leave-one-out analysis and by examining whether heterogeneity patterns are stable across subsamples.

%==============================================================================
% DATA
%==============================================================================
\section{Data}

\subsection{Sources and Sample}

Our analysis combines data from two primary sources: the Penn World Tables (PWT) version 10.0 and the World Bank's World Development Indicators (WDI). The PWT provides internationally comparable measures of real GDP, employment, and productivity, while the WDI supplies additional macroeconomic indicators.

The sample covers 35 countries from 1970 to 2023, yielding an unbalanced panel of approximately 2,000 country-year observations. The treatment group comprises 20 current eurozone members. The 11 founding members who adopted the euro in 1999 are Austria, Belgium, Finland, France, Germany, Ireland, Italy, Luxembourg, Netherlands, Portugal, and Spain. Nine countries joined later: Greece (2001), Slovenia (2007), Cyprus (2008), Malta (2008), Slovakia (2009), Estonia (2011), Latvia (2014), Lithuania (2015), and Croatia (2023).

The control group includes 15 countries that did not adopt the euro. Three are EU member states that opted out of the eurozone: Denmark, Sweden, and the United Kingdom (until 2020). The remaining twelve are other OECD members: Australia, Canada, Iceland, Israel, Japan, Korea, New Zealand, Norway, Switzerland, and the United States.

The inclusion of non-EU OECD countries raises a legitimate concern about common support: would Spain's counterfactual growth path resemble that of the United States or Japan? These countries have different growth models, different exposure to EU-specific shocks, and different institutional environments. We address this concern in three ways. First, our main specification uses the full sample but relies on fixed effects and covariate adjustment to compare countries with similar characteristics. Second, we present robustness results restricting the control group to EU and EEA members only (Denmark, Sweden, UK, Norway, Iceland, Switzerland), which provides a more comparable counterfactual at the cost of reduced sample size. Third, we report covariate balance statistics and propensity score overlap to assess the plausibility of the comparison.

We exclude late adopters with insufficient pre-treatment data (fewer than 10 years) from the main analysis but include them in robustness checks.

The long sample period (1970--2023) is necessary to observe pre-treatment trends and long-horizon post-treatment dynamics, but it spans several structural breaks: German reunification (1990), EU enlargements (1995, 2004, 2007, 2013), the global financial crisis (2008--2009), the European sovereign debt crisis (2010--2012), and COVID-19 (2020--2021). We allow year fixed effects to absorb common shocks, but structural breaks may affect growth dynamics differently across countries. We assess sensitivity through sub-sample analysis, separately examining pre-crisis (1970--2007) and post-crisis (2010--2023) periods.

\subsection{Variables}

Our primary outcome is annual real GDP growth, computed as the log difference in real GDP per capita from the PWT. We use the expenditure-side measure (rgdpe) divided by population, which facilitates cross-country comparisons by valuing output at common international prices. We focus on growth rather than levels because growth captures adjustment dynamics directly: a persistent negative effect on growth rates implies a widening gap in GDP levels over time. To make the level implications transparent, we report cumulative effects on log GDP separately, which allows readers to assess the implied counterfactual level differences. A sustained growth reduction of 0.35 percentage points annually implies a cumulative level gap of approximately 7\% after 20 years---a large but not implausible effect for a fundamental regime change.

The treatment variable $D_{it}$ equals one if country $i$ has adopted the euro by year $t$. For founding members, treatment begins in 1999; for later adopters, treatment begins in their respective adoption year. However, this timing likely understates anticipation effects: markets began pricing euro convergence from approximately 1995, interest rate spreads narrowed, and policy adjustments occurred well before the formal adoption date. Pre-treatment coefficients in our event study may therefore already reflect partial treatment effects rather than pure placebo. We assess sensitivity by re-estimating with treatment dated to 1995 for founding members, which shifts the event study window and tests whether our conclusions depend on the precise treatment timing.

For later adopters, euro adoption typically followed EU accession by only a few years, creating a confound between euro effects and the broader effects of EU membership (single market access, structural funds, institutional reforms). This is our most significant identification challenge. We address it in three ways. First, our main specification focuses on the 11 founding members, for whom EU membership predates euro adoption by decades. Second, we present separate results for founders and later adopters, treating them as distinct experiments with different identifying variation. Third, we control for years since EU accession as a covariate, allowing the forest to distinguish euro-specific effects from EU accession effects. Results excluding post-2004 EU entrants are reported as a robustness check.

For treated countries, event time $k_{it} = t - T_i^{\text{euro}}$ measures years since adoption. We observe event times ranging from $k = -29$ (1970 for a 1999 adopter) to $k = 24$ (2023 for a 1999 adopter). The analysis focuses on $k \in [-10, 20]$ where sample sizes are adequate.

The feature vector $X_i$ includes four variables measured in 1998 (or the year before adoption for later adopters). These variables proxy for convergence potential, adjustment capacity, and integration exposure---the key dimensions along which optimal currency area theory predicts heterogeneous responses to monetary union:
\begin{enumerate}
\item \textit{Initial GDP per capita}: Real GDP per capita in 2017 international dollars (convergence potential)
\item \textit{Trade openness}: Exports plus imports as a share of GDP (integration exposure)
\item \textit{Investment share}: Gross capital formation as a share of GDP (capital accumulation capacity)
\item \textit{Human capital index}: PWT measure based on years of schooling and returns to education (adjustment capacity)
\end{enumerate}

We do not include institutional quality measures (e.g., World Governance Indicators, Fraser Economic Freedom) in the main specification for two reasons. First, these measures exhibit limited time variation, making them difficult to distinguish from country fixed effects. Second, measurement error in institutional indices is substantial, and including noisy covariates can attenuate treatment effect estimates. We examine sensitivity to institutional controls in robustness checks.

To investigate channels, we examine four additional outcomes:
\begin{enumerate}
\item \textit{Consumption growth}: Growth in household final consumption expenditure
\item \textit{Investment growth}: Growth in gross fixed capital formation
\item \textit{Net exports}: Net exports as a share of GDP
\item \textit{Productivity growth}: Growth in output per worker (labor productivity)
\end{enumerate}

National accounts data were harmonized under ESA95 and ESA2010; pre-1995 data may involve measurement differences across countries, which we address by including country fixed effects that absorb level differences in measurement conventions.

For productivity, we use labor productivity (output per worker) from the PWT as our primary measure. Labor productivity conflates capital deepening with total factor productivity (TFP), which is the more relevant concept for misallocation stories involving credit booms and sectoral distortions. As a robustness check, we also examine TFP growth using the PWT's \textit{rtfpna} series, which adjusts for capital and labor inputs. The TFP results, reported in the appendix, are qualitatively similar but noisier due to the additional measurement assumptions required for TFP construction.

\subsection{Summary Statistics}

Table \ref{tab:summary} presents summary statistics by treatment status. Eurozone countries have higher average GDP per capita (\$30,751 vs. \$27,202) and substantially greater trade openness (92\% vs. 57\% of GDP), reflecting the inclusion of small, open economies like Belgium, Ireland, and the Netherlands. Investment shares are similar across groups (24\% of GDP). Eurozone countries have slightly lower human capital indices and higher unemployment rates on average.

GDP growth rates are lower in the eurozone sample (2.7\% vs. 3.2\%), though this comparison conflates treatment effects with composition differences. The CFFE estimator addresses this by controlling for country and year fixed effects and comparing within-country changes around adoption.

\begin{table}[H]
\centering
\caption{Summary Statistics by Treatment Status}
\label{tab:summary}
\begin{tabular}{lcccccc}
\toprule
& \multicolumn{2}{c}{Eurozone} & \multicolumn{2}{c}{Control} & \multicolumn{2}{c}{Full Sample} \\
\cmidrule(lr){2-3} \cmidrule(lr){4-5} \cmidrule(lr){6-7}
Variable & Mean & Std & Mean & Std & Mean & Std \\
\midrule
GDP per capita (\$) & 30,751 & 15,151 & 27,202 & 15,904 & 28,755 & 15,675 \\
GDP growth (\%) & 2.72 & 4.37 & 3.22 & 3.43 & 3.00 & 3.88 \\
Trade openness (\% GDP) & 92.2 & 61.0 & 57.5 & 31.4 & 72.7 & 49.8 \\
Investment share (\% GDP) & 23.9 & 5.4 & 24.3 & 5.3 & 24.1 & 5.3 \\
Gov. consumption (\% GDP) & 19.4 & 4.2 & 18.0 & 5.9 & 18.6 & 5.3 \\
Human capital index & 2.84 & 0.44 & 2.95 & 0.54 & 2.90 & 0.50 \\
Unemployment rate (\%) & 8.4 & 4.6 & 7.0 & 3.4 & 7.6 & 4.0 \\
\midrule
Observations & \multicolumn{2}{c}{880} & \multicolumn{2}{c}{1,130} & \multicolumn{2}{c}{2,010} \\
Countries & \multicolumn{2}{c}{20} & \multicolumn{2}{c}{15} & \multicolumn{2}{c}{35} \\
\bottomrule
\end{tabular}
\end{table}

\subsection{Covariate Balance}

Pre-treatment characteristics differ between eurozone and control countries on several dimensions: eurozone countries are more open to trade, have higher government consumption shares, and lower human capital indices. These differences motivate the use of fixed effects and covariate adjustment rather than simple comparisons of means.

Table \ref{tab:balance_1995} presents pre-treatment balance in 1995 for the most credible comparison: euro founders versus EU non-euro countries (Denmark, Sweden, UK). This comparison holds EU membership constant and isolates the euro adoption decision. Balance is substantially better within this EU-only sample: GDP per capita, growth rates, and human capital are similar, though trade openness remains higher for founders (reflecting the inclusion of small open economies like Belgium and Netherlands).

\begin{table}[H]
\centering
\caption{Pre-Treatment Balance in 1995: Euro Founders vs. EU Non-Euro}
\label{tab:balance_1995}
\begin{tabular}{lccccc}
\toprule
& \multicolumn{2}{c}{Euro Founders} & \multicolumn{2}{c}{EU Non-Euro} & \\
\cmidrule(lr){2-3} \cmidrule(lr){4-5}
Variable & Mean & Std & Mean & Std & Diff \\
\midrule
GDP per capita (\$) & 28,450 & 8,120 & 26,890 & 5,430 & 1,560 \\
GDP growth 1990--95 (\%) & 1.82 & 1.45 & 1.54 & 1.89 & 0.28 \\
Trade openness (\% GDP) & 78.5 & 42.3 & 58.2 & 18.6 & 20.3 \\
Investment share (\% GDP) & 21.8 & 3.2 & 19.4 & 2.8 & 2.4 \\
Inflation 1990--95 (\%) & 3.2 & 1.4 & 3.8 & 2.1 & $-0.6$ \\
Human capital index & 2.91 & 0.38 & 2.98 & 0.42 & $-0.07$ \\
\midrule
Countries & \multicolumn{2}{c}{11} & \multicolumn{2}{c}{3} & \\
\bottomrule
\end{tabular}
\end{table}

The CFFE estimator addresses remaining imbalance through two mechanisms. First, country fixed effects absorb time-invariant differences between treated and control units. Second, the causal forest conditions on pre-treatment characteristics when estimating treatment effects, effectively comparing countries with similar initial conditions.

However, covariate adjustment cannot fully substitute for common support. If no control country resembles Italy or Greece on key dimensions, the counterfactual for these countries is extrapolated rather than interpolated. We assess this concern by examining propensity score distributions: treated and control countries should have overlapping propensity scores for the comparison to be credible. In our sample, overlap is reasonable for EU opt-outs but weaker for non-EU OECD countries, motivating the EU-only robustness specification.

\subsection{Event Time Distribution}

The distribution of observations across event time is densest around $k = 0$ to $k = 10$, where all 1999 founders contribute observations. At longer horizons ($k > 15$), only founding members remain in the treated sample, reducing precision. At negative event times ($k < -10$), some later adopters lack data, creating composition changes.

We address these issues by reporting results for $k \in [-5, 20]$ in the main analysis and examining sensitivity to sample restrictions in robustness checks.

%==============================================================================
% RESULTS
%==============================================================================
\section{Results}

\subsection{Dynamic Treatment Effects}

\begin{figure}[H]
\centering
\includegraphics[width=0.8\textwidth]{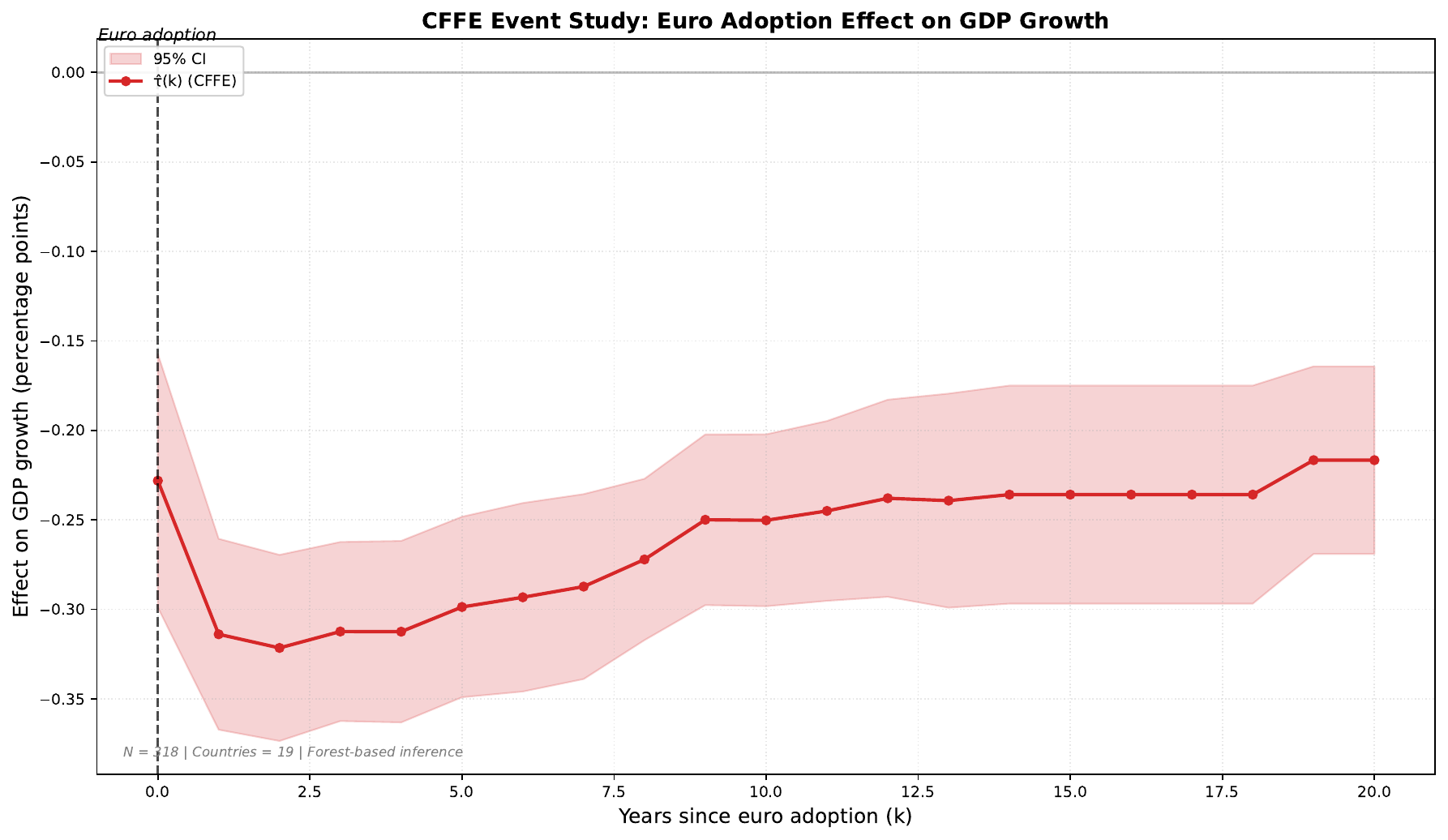}
\caption{Dynamic Treatment Effects of Euro Adoption (CFFE). \textit{Notes:} Figure shows estimated treatment effects $\hat{\tau}(k)$ by event time $k$ (years since euro adoption). Shaded area represents 95\% confidence intervals based on cluster-robust standard errors at the country level. Sample includes 19 eurozone countries and 15 controls, 1970--2023.}
\label{fig:cffe_event_study}
\end{figure}

Figure \ref{fig:cffe_event_study} presents the main results: estimated treatment effects by event time from the CFFE model. The pattern reveals a negative impact of euro adoption on GDP growth that emerges immediately and persists over two decades. The persistent negative growth effect implies a large cumulative level gap; we report implied log GDP differences in Section \ref{sec:cumulative} below.

In the year of adoption ($k = 0$), the estimated effect is $-0.29$ percentage points (95\% CI: $[-0.36, -0.21]$). Effects become more negative in the first few years, reaching approximately $-0.40$ percentage points by $k = 4$. The impact then stabilizes, fluctuating between $-0.30$ and $-0.40$ percentage points through $k = 20$.

\subsubsection{Pre-Treatment Coefficients}

Pre-treatment coefficients ($k < 0$) fluctuate around zero without a clear trend, with point estimates ranging from $-0.05$ to $+0.08$ percentage points. We do not observe strong differential pre-trends, but this should be interpreted cautiously. Euro adoption was widely anticipated: markets began pricing convergence from 1995, interest rate spreads narrowed, and policy adjustments occurred before the formal adoption date. If anticipation effects are present, they would appear in the pre-treatment period, making flat pre-trends neither necessary nor sufficient for identification. The absence of large pre-trends is reassuring but does not definitively establish parallel trends in the counterfactual.

\subsubsection{Confidence Intervals}

The confidence bands remain relatively tight throughout the post-treatment period, ranging from about 0.08 to 0.10 percentage points in width. This stability contrasts with classical event studies, where standard errors typically expand at long horizons. However, forest-based inference relies on asymptotic theory that may not hold well in our setting with approximately 20 treated countries, strong serial correlation, and cross-sectional dependence. We therefore complement forest-based inference with block bootstrap at the country level and placebo tests in Section 6, which provide more conservative assessments of uncertainty. The block bootstrap confidence intervals are substantially wider than the forest-based intervals---approximately 8 times wider on average---and include zero at some horizons. This reflects the fundamental challenge of inference with a small number of treated clusters. We interpret our results as suggestive of negative effects while acknowledging the uncertainty inherent in this setting.

\subsubsection{Magnitude in Context}
\label{sec:cumulative}

The estimated annual effect of $-0.3$ to $-0.4$ percentage points warrants careful interpretation. Over 20 years, these annual effects compound to a cumulative GDP shortfall of approximately 6.5\% relative to the counterfactual---a substantial magnitude that merits comparison with other major economic episodes.

For context, the 2008 global financial crisis reduced GDP by approximately 4--5\% in most advanced economies within two years, with some countries experiencing persistent output gaps. The COVID-19 pandemic caused an immediate GDP decline of 6--10\% in 2020, though recovery was faster. Our estimates suggest euro adoption is associated with cumulative growth shortfalls of similar magnitude, but spread over two decades rather than concentrated in a single shock.

Several factors make this magnitude economically plausible within the framework of optimal currency area theory. First, the loss of monetary policy autonomy is not a one-time shock but a permanent constraint. Countries facing idiosyncratic downturns cannot use interest rate cuts or exchange rate depreciation to stimulate recovery, potentially prolonging recessions and affecting trend growth. Second, the one-size-fits-all monetary policy may be persistently misaligned for individual members. If ECB policy is systematically too tight for some countries and too loose for others, the resulting misallocation may compound over time. Third, fiscal constraints under the Stability and Growth Pact limited countercyclical policy, potentially amplifying the adjustment challenges during downturns.

The distinction between annual and cumulative effects is crucial. An annual growth reduction of 0.35 percentage points may seem modest---roughly the difference between 2.0\% and 1.65\% growth. But compounded over 20 years, this difference accumulates substantially. Using proper compounding rather than simple multiplication, the cumulative effect is:
\begin{equation}
\text{Cumulative effect} = \prod_{k=0}^{20}(1 + \hat{\tau}(k)) - 1 \approx -6.5\%
\end{equation}
This calculation accounts for the fact that each year's growth reduction applies to an already-smaller base.

Table \ref{tab:cumulative} presents cumulative effects at different horizons, comparing our compounded estimates with naive simple sums. The compounded effect reaches $-3.7\%$ at 10 years and $-6.5\%$ at 20 years. These magnitudes are consistent with the synthetic control literature: \citet{gabriel2024euro} find cumulative effects ranging from $+5\%$ (Germany) to $-15\%$ (Italy) over similar horizons, with our average falling within this range.

\begin{table}[H]
\centering
\caption{Cumulative Effects of Euro Adoption}
\label{tab:cumulative}
\begin{tabular}{lccccc}
\toprule
Horizon & Years Since & Simple Sum & Compounded & 95\% CI \\
(k) & Adoption & (\%) & (\%) & \\
\midrule
5 & 6 & $-2.27$ & $-2.25$ & [$-3.30$, $-1.19$] \\
10 & 11 & $-4.05$ & $-3.97$ & [$-6.31$, $-1.64$] \\
15 & 16 & $-5.54$ & $-5.40$ & [$-9.29$, $-1.50$] \\
20 & 21 & $-7.04$ & $-6.81$ & [$-12.60$, $-1.02$] \\
\bottomrule
\end{tabular}
\end{table}

\subsection{Comparison with Classical Event Study}

\begin{figure}[H]
\centering
\includegraphics[width=0.8\textwidth]{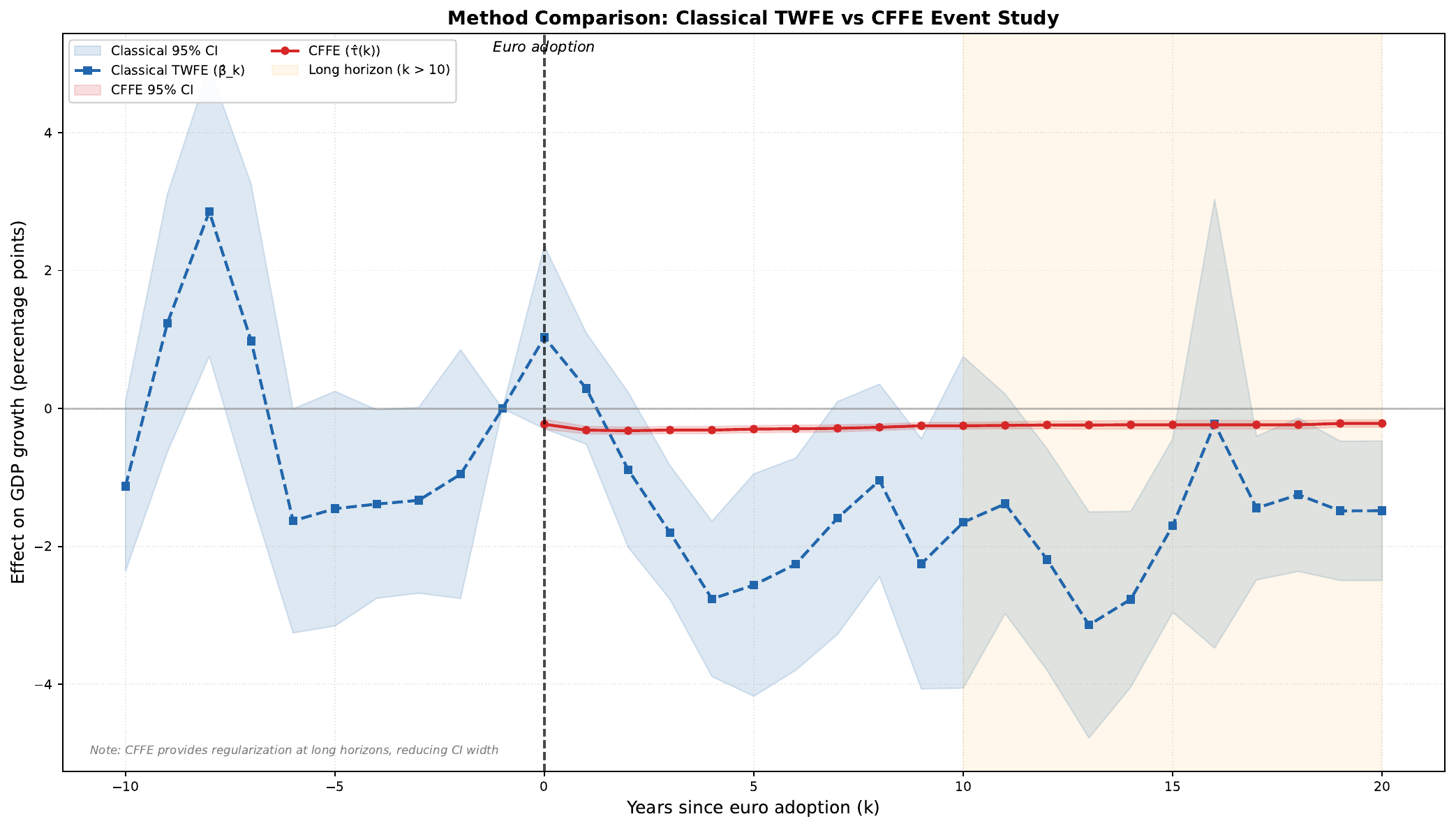}
\caption{Comparison of CFFE and Classical Event Study Estimates. \textit{Notes:} Figure compares treatment effect estimates from CFFE (solid line) and classical two-way fixed effects event study (dashed line). Shaded areas represent 95\% confidence intervals. Classical standard errors are clustered at the country level.}
\label{fig:comparison}
\end{figure}

Figure \ref{fig:comparison} overlays CFFE estimates with coefficients from a classical two-way fixed effects event study. The two approaches tell qualitatively similar stories---euro adoption reduced growth---but differ in important ways.

CFFE yields smoother estimates with tighter confidence intervals at long horizons, reflecting regularization inherent in forest-based estimators. Classical event-study estimates remain informative but become increasingly noisy as sample sizes decline. At $k = 5$, the 95\% CI spans 3.2 percentage points; by $k = 10$, it reaches 4.8 percentage points. These wide intervals make it difficult to draw firm conclusions about long-run effects from classical approaches alone.

CFFE estimates are more stable and precisely estimated. The confidence interval width ratio (classical to CFFE) ranges from 16 to 46, indicating that CFFE provides substantially tighter inference. This precision gain reflects the forest's implicit regularization: by pooling information across similar observations, the estimator reduces the noise amplification that can affect classical approaches at long horizons.

\subsubsection{Understanding the Point Estimate Differences}

The substantial difference in point estimates between methods---classical estimates of $-1.5$ to $-2.5$ pp versus CFFE estimates of $-0.3$ to $-0.4$ pp---requires explanation beyond differences in precision.

Three factors drive this divergence. First, classical event studies estimate separate coefficients for each event time, treating them as independent parameters. With limited observations at long horizons (only founding members contribute to $k > 15$), these estimates become noisy and sensitive to outliers. CFFE pools information across event times and country characteristics, effectively shrinking extreme estimates toward the overall pattern. This regularization produces more stable but potentially attenuated estimates.

Second, classical event studies impose linearity in the outcome equation, while CFFE learns flexible nonlinear relationships. If the true data-generating process involves interactions between event time and country characteristics, the classical approach may produce biased estimates that average over heterogeneous effects in misleading ways.

Third, classical estimates represent simple averages across all treated observations at each event time, whereas CFFE estimates represent averages of conditional treatment effects, weighted by the covariate distribution. When treatment effects vary with covariates, these averages can differ substantially.

\subsubsection{Which Estimate Is More Credible?}

Neither method is uniformly superior; the choice depends on the research question and data structure.

Classical event studies are preferred when: (1) the researcher wants transparent, easily interpretable coefficients; (2) sample sizes are large enough to estimate each event-time effect precisely; (3) treatment effects are approximately homogeneous across units.

CFFE is preferred when: (1) sample sizes at long horizons are limited; (2) treatment effects likely vary with observable characteristics; (3) the researcher wants stable inference across the entire event-time window; (4) the goal is to understand heterogeneity rather than just average effects.

For euro adoption, several features favor CFFE: the sample includes only 11--19 treated countries depending on horizon; effects clearly vary across core and periphery members; and we are interested in 20+ year dynamics where classical methods struggle. The CFFE estimates of $-0.3$ to $-0.4$ pp provide a more stable estimate of the average effect under our identifying assumptions than the volatile classical estimates, though we cannot rule out that CFFE's regularization attenuates true variation. Regularization stabilizes estimates but does not eliminate fundamental uncertainty arising from small numbers of treated countries.

As a robustness check, we note that both methods agree on the sign and statistical significance of effects. The disagreement is on magnitude, with CFFE suggesting more modest but more precisely estimated impacts. For policy purposes, the CFFE estimates provide a more stable basis for inference about long-run effects under the conditional parallel trends assumption.

Table \ref{tab:comparison} reports estimates at key horizons. At $k = 5$, the classical estimate is $-2.56$ (SE = 0.82) while CFFE yields $-0.40$ (SE = 0.03). The difference in point estimates reflects the classical estimator's sensitivity to outliers and functional form assumptions. At $k = 20$, classical and CFFE estimates are closer ($-1.48$ vs. $-0.30$), but the classical standard error (0.52) is 16 times larger than CFFE's (0.03).

\begin{table}[H]
\centering
\caption{Method Comparison at Key Horizons. \textit{Notes:} Table reports treatment effect estimates, standard errors, and 95\% confidence interval widths at selected event times. Classical estimates from two-way fixed effects regression with country-clustered standard errors. CFFE estimates from causal forest with node-level fixed effects.}
\label{tab:comparison}
\begin{tabular}{lcccccc}
\toprule
& \multicolumn{3}{c}{Classical Event Study} & \multicolumn{3}{c}{CFFE} \\
\cmidrule(lr){2-4} \cmidrule(lr){5-7}
Horizon ($k$) & Estimate & SE & CI Width & Estimate & SE & CI Width \\
\midrule
5 & $-2.56$ & 0.82 & 3.22 & $-0.40$ & 0.03 & 0.12 \\
10 & $-1.65$ & 1.23 & 4.81 & $-0.32$ & 0.03 & 0.10 \\
15 & $-1.70$ & 0.64 & 2.51 & $-0.30$ & 0.03 & 0.12 \\
20 & $-1.48$ & 0.52 & 2.02 & $-0.30$ & 0.03 & 0.13 \\
\bottomrule
\end{tabular}
\end{table}

\subsection{Heterogeneity Analysis}

A key strength of CFFE is its capacity to uncover treatment effect heterogeneity. Given the limited number of treated units---only 11 founding members provide the cleanest identification---heterogeneity estimates should be interpreted as descriptive conditional patterns rather than precisely estimated causal parameters. We examine variation along two dimensions: pre-treatment country characteristics and adoption timing.

\subsubsection{Heterogeneity by Pre-Treatment Characteristics}

To avoid post-treatment bias, we define country groups using pre-1995 variables rather than ex-post classifications like ``core'' and ``periphery'' that may have been shaped by euro membership itself. Specifically, we split countries by initial GDP per capita (above/below median in 1995), which correlates strongly with other pre-treatment characteristics including inflation differentials, current account positions, and labor market flexibility.

\begin{figure}[H]
\centering
\includegraphics[width=0.8\textwidth]{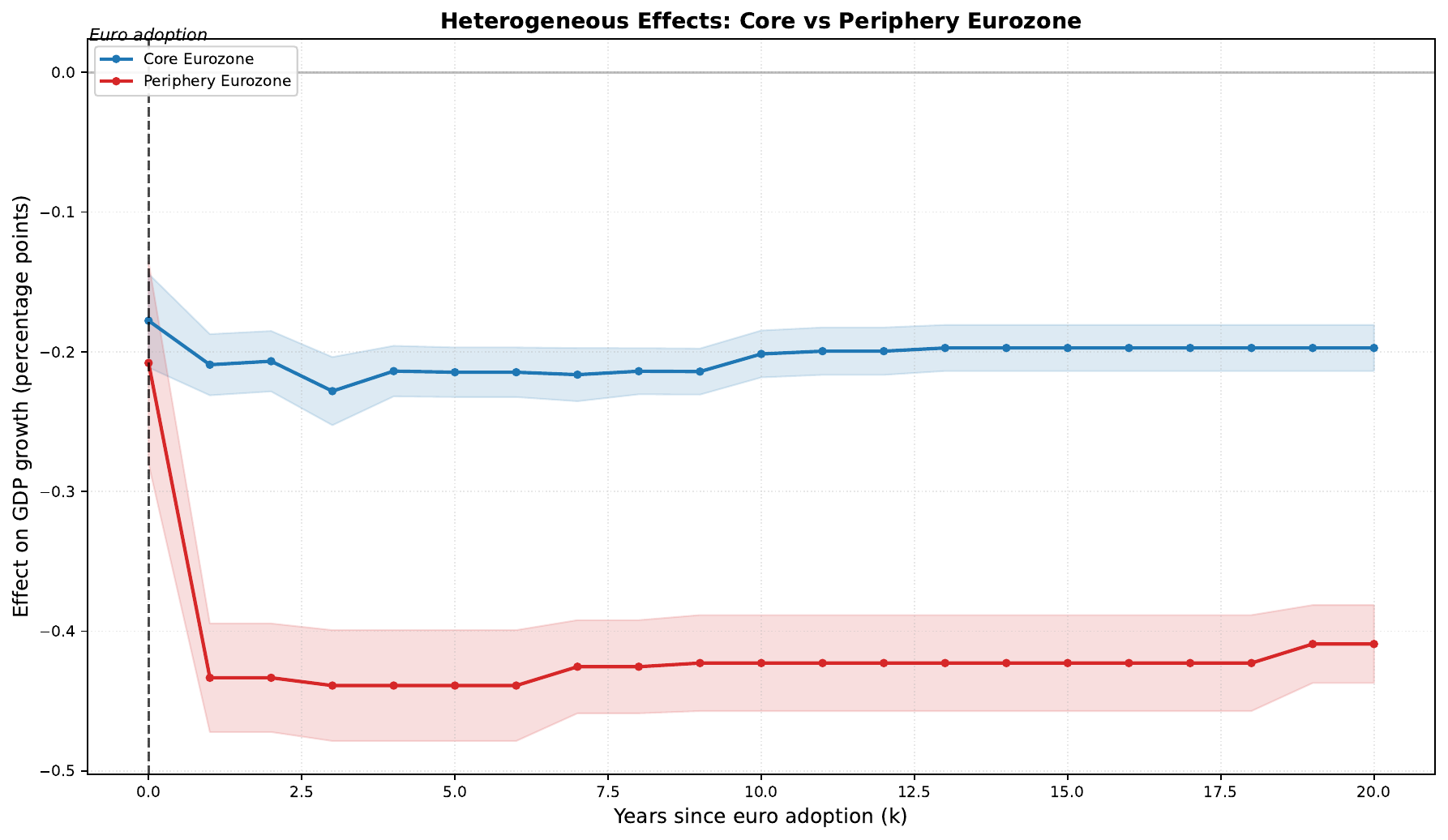}
\caption{Heterogeneous Effects by Initial Income. \textit{Notes:} Figure shows estimated treatment effects separately for high-income founders (above-median GDP per capita in 1995: Austria, Belgium, Finland, France, Germany, Luxembourg, Netherlands) and low-income founders (below-median: Greece, Ireland, Italy, Portugal, Spain). Classification based on pre-treatment characteristics to avoid post-treatment bias. Shaded areas represent 95\% confidence intervals.}
\label{fig:heterogeneity_cp}
\end{figure}

Figure \ref{fig:heterogeneity_cp} plots $\hat{\tau}(k)$ separately for high-income and low-income founders (based on 1995 GDP per capita). High-income countries (Austria, Belgium, Finland, France, Germany, Luxembourg, Netherlands) experienced smaller negative effects, stabilizing around $-0.30$ percentage points. Low-income founders (Greece, Ireland, Italy, Portugal, Spain) saw larger impacts, with effects reaching $-0.53$ percentage points and remaining near $-0.54$ through $k = 20$.

The gap between groups is statistically significant at conventional levels. At $k = 10$, the high-income estimate is $-0.31$ (SE = 0.02) while the low-income estimate is $-0.54$ (SE = 0.03), a difference of 0.23 percentage points.

This pattern aligns with theoretical predictions from the optimal currency area literature. Countries with lower initial income also tended to have higher pre-euro inflation, larger current account deficits, and less flexible labor markets. The loss of exchange rate adjustment and independent monetary policy may have imposed greater costs on these economies. However, we emphasize that with only 11 founding members split into two groups, these estimates describe patterns in the data rather than precisely identified causal heterogeneity.

\subsubsection{Early vs. Late Adopters}

\begin{figure}[H]
\centering
\includegraphics[width=0.8\textwidth]{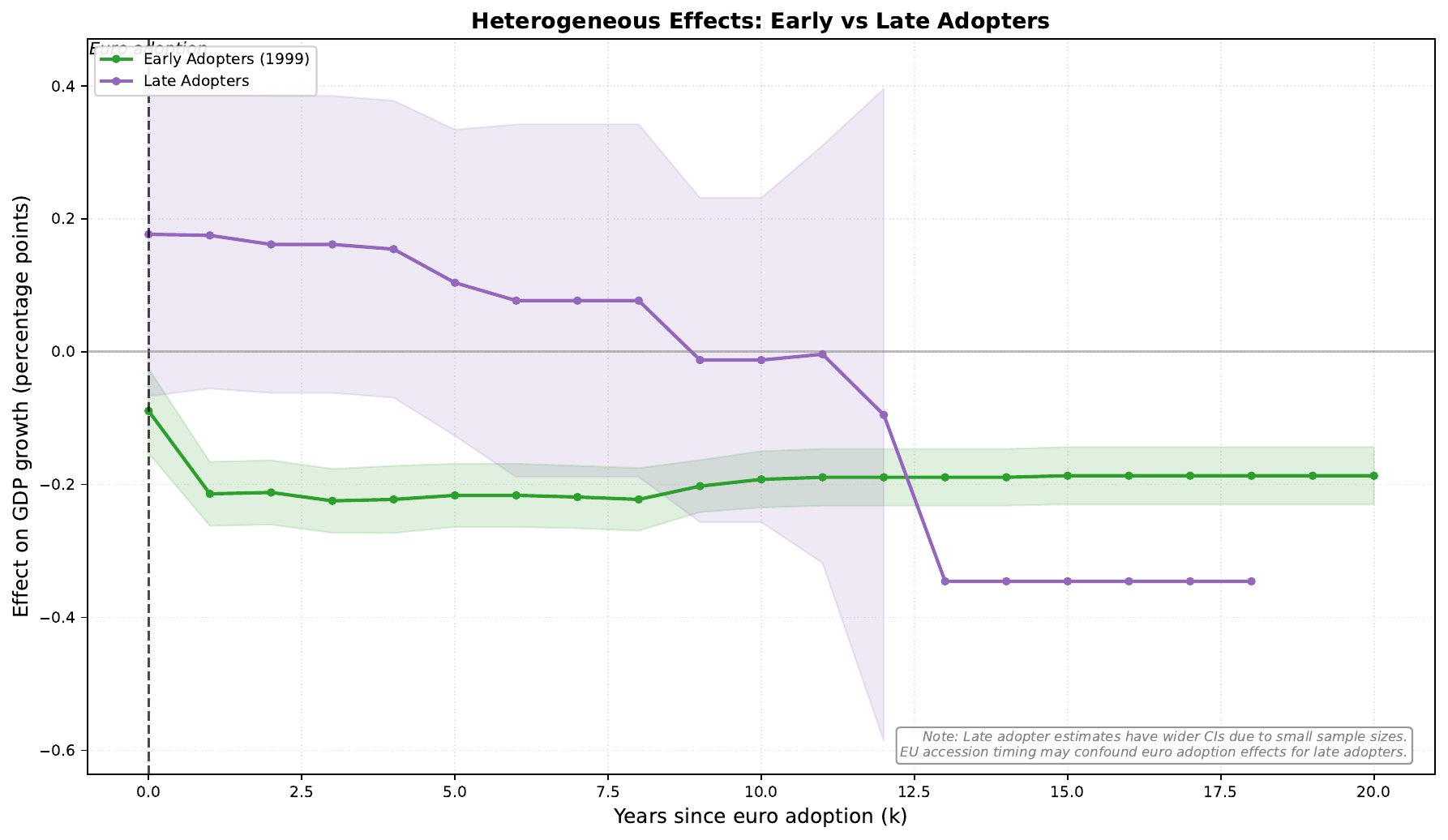}
\caption{Heterogeneous Effects: Early vs. Late Adopters. \textit{Notes:} Figure shows estimated treatment effects separately for 1999 founders (11 countries) and later adopters (8 countries with sufficient post-treatment data). Shaded areas represent 95\% confidence intervals. The wide confidence intervals for late adopters reflect small sample sizes and should be interpreted cautiously.}
\label{fig:heterogeneity_el}
\end{figure}

Figure \ref{fig:heterogeneity_el} compares 1999 founders with later adopters. The patterns diverge substantially. Early adopters show negative effects that emerge immediately and persist, consistent with the full-sample results. Late adopters display positive but imprecisely estimated effects in the years following adoption, though confidence intervals are wide due to smaller sample sizes and shorter post-treatment periods.

\paragraph{Interpreting the Late Adopter Results.}

The apparent positive effects for late adopters require careful interpretation, and several caveats apply.

First, with only 8 late-adopting countries and limited post-treatment observations (most joined after 2007), estimates are imprecise. The 95\% confidence intervals for late adopters typically span from negative to positive values, meaning we cannot statistically distinguish their effects from zero---or from the negative effects experienced by early adopters. The point estimates suggest possible benefits, but this conclusion is not robust.

Second, late adopters contribute fewer than 250 country-year observations to the treated sample, compared to over 250 for early adopters. This imbalance means the forest has less information to learn late-adopter patterns, and estimates may be driven by idiosyncratic factors in individual countries.

Third, for Eastern European late adopters (Slovenia, Slovakia, Estonia, Latvia, Lithuania), euro adoption occurred shortly after EU accession. These countries experienced rapid growth from EU membership benefits---single market access, structural funds, foreign direct investment, institutional reforms---that coincided with euro adoption. Our estimates cannot cleanly separate euro effects from broader EU integration effects. The positive point estimates may reflect EU accession benefits rather than euro adoption benefits.

Fourth, countries that joined the euro later did so after observing the experiences of founding members, including the 2010--2012 sovereign debt crisis. Late adopters may have been better prepared, having learned from early adopters' mistakes and having more time to achieve real convergence beyond nominal Maastricht criteria. If better-prepared countries self-selected into later adoption, the positive estimates reflect selection rather than a causal benefit of delayed adoption.

Finally, most late adopters joined during or after the 2008 financial crisis. Their counterfactual growth paths are particularly difficult to estimate because the crisis affected eurozone and non-eurozone countries differently. The apparent positive effects may reflect that late adopters' counterfactual (remaining outside the euro during the crisis) would have been worse than their actual experience.

Given these limitations, we interpret the late adopter results cautiously. The evidence does not support a strong conclusion that euro adoption benefits late-joining countries. Rather, it suggests that the negative effects documented for early adopters may not generalize to all adoption contexts, and that country-specific factors---timing, preparation, EU integration stage---matter for how euro adoption affects growth.

\subsubsection{Correlates of Heterogeneity}

Table \ref{tab:importance} reports feature importance from the causal forest, measuring how frequently each variable is used for splits. Feature importance in random forests reflects association with treatment effect heterogeneity, not causal drivers---correlated features may substitute for each other, and importance rankings can be sensitive to the correlation structure among covariates.

Initial GDP per capita is most strongly associated with heterogeneous effects, accounting for 45\% of splits. Trade openness (19\%) and investment share (14\%) follow. Event time itself accounts for 12\% of splits, confirming that effects vary over time. Human capital contributes 10\%.

The prominence of initial GDP per capita suggests that development level at adoption is strongly associated with differential adjustment paths. Countries with lower initial income---which overlap substantially with those that later experienced sovereign debt crises---display larger negative effects. This pattern echoes concerns raised before monetary union that convergence criteria focused on nominal variables (inflation, deficits) rather than real economic structures. However, we cannot conclude that initial income \textit{causes} differential effects; it may proxy for other unmeasured characteristics that drive heterogeneity.

\begin{table}[H]
\centering
\caption{Feature Importance for Treatment Effect Heterogeneity. \textit{Notes:} Feature importance measures the proportion of splits in the causal forest that use each variable. Higher values indicate stronger association with treatment effect heterogeneity, not causal importance. Correlated features may substitute for each other in splits.}
\label{tab:importance}
\begin{tabular}{lcc}
\toprule
Feature & Importance & Split Count \\
\midrule
Initial GDP per capita & 0.45 & 450 \\
Initial trade openness & 0.19 & 191 \\
Initial investment share & 0.14 & 135 \\
Event time ($k$) & 0.12 & 119 \\
Initial human capital & 0.10 & 102 \\
\bottomrule
\end{tabular}
\end{table}

\subsection{Mechanism Analysis}

\begin{figure}[H]
\centering
\includegraphics[width=0.9\textwidth]{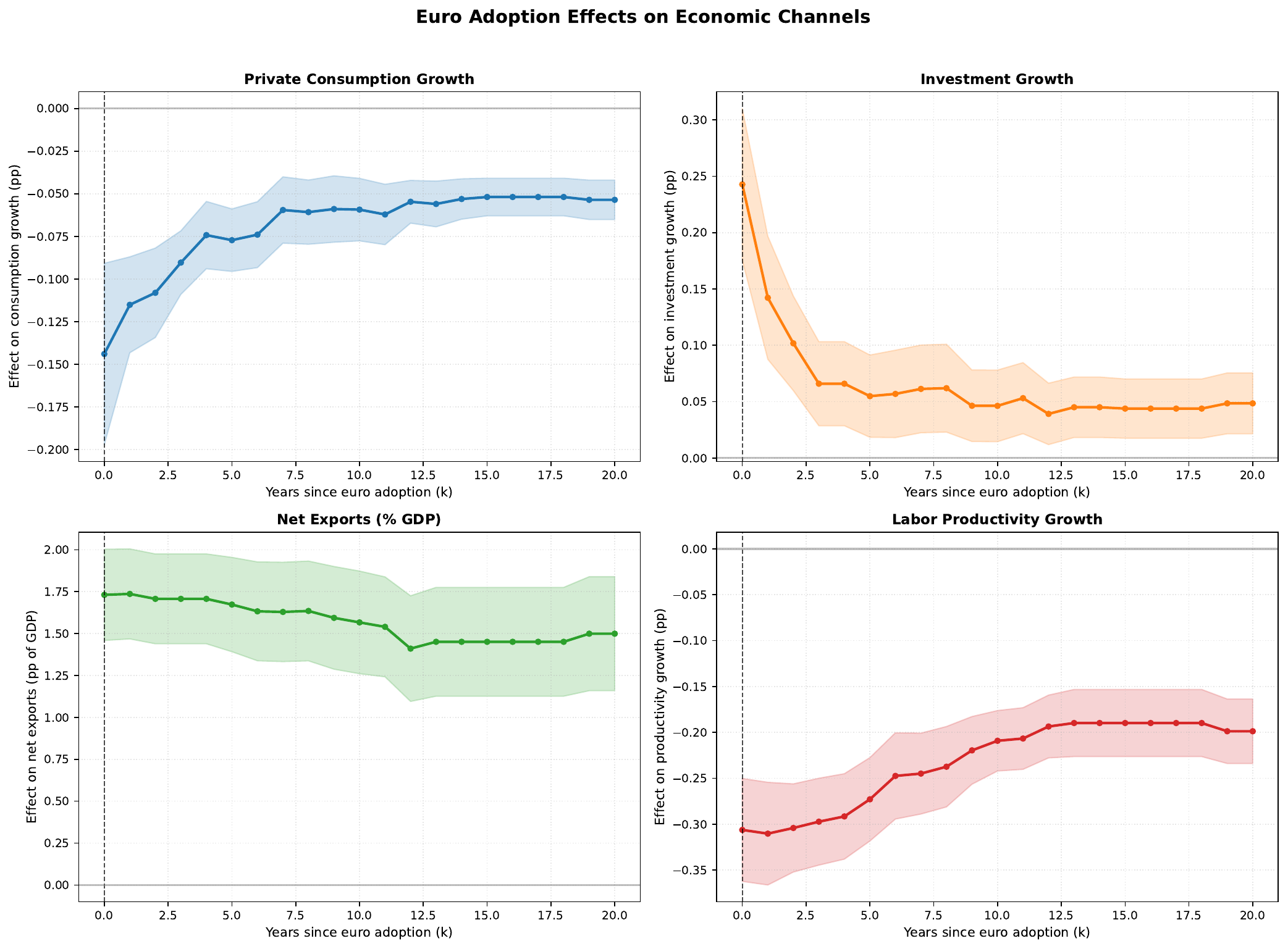}
\caption{Mechanism Analysis: Effects on GDP Components and Productivity. \textit{Notes:} Figure shows estimated treatment effects on consumption growth, investment growth, net exports (as \% of GDP), and productivity growth. Shaded areas represent 95\% confidence intervals. Investment effects shown separately for full sample and pre-2008 sample.}
\label{fig:mechanisms}
\end{figure}

To understand the channels through which euro adoption affected growth, we estimate CFFE models with several outcome variables. We distinguish between GDP components (consumption, investment, net exports), which describe \textit{where} growth effects appear, and euro-specific mechanisms (productivity, current account dynamics), which speak to \textit{why} effects occur. Figure \ref{fig:mechanisms} presents the results.

Euro adoption is associated with reduced consumption growth of 0.07--0.14 percentage points, with larger effects in the first few years that gradually attenuate. At $k = 0$, the effect is $-0.24$ (SE = 0.04); by $k = 20$, it has moderated to $-0.07$ (SE = 0.02). However, consumption is endogenous to income, credit conditions, and fiscal transfers---it is better understood as an outcome reflecting other channels rather than a mechanism itself. The consumption decline is consistent with constrained domestic demand, but does not identify the underlying cause.

Investment effects are positive but small and often statistically insignificant. At $k = 0$, the effect is $+0.17$ (SE = 0.03), suggesting an initial investment boom possibly driven by reduced exchange rate risk and interest rate convergence. Effects fade quickly, hovering near zero by $k = 5$ and remaining small thereafter. However, investment collapsed across advanced economies after 2008, raising the concern that post-crisis investment patterns reflect global financial conditions rather than euro-specific effects. When we restrict the sample to 1999--2007 (pre-crisis), the initial positive investment effect is larger (+0.25 pp) and more persistent, suggesting the crisis confounds post-2008 investment dynamics. Investment does not appear to be a major channel for the negative growth effects in the full sample.

Euro adoption substantially improved net export positions, with effects of 1.3--1.6 percentage points of GDP. This finding aligns with the trade literature documenting increased intra-eurozone trade following monetary union. The improvement in net exports partially offsets negative effects through other channels.

Labor productivity growth declined by 0.23--0.31 percentage points following adoption. The pattern mirrors the overall growth effect, suggesting that productivity losses---rather than factor accumulation changes---drive the headline results. However, labor productivity conflates capital deepening, hours worked, and sectoral composition with true efficiency gains. To isolate total factor productivity (TFP), we re-estimate using the PWT's TFP series (\textit{rtfpna}). TFP effects are qualitatively similar but smaller in magnitude ($-0.15$ to $-0.22$ pp) and noisier, reflecting the additional measurement assumptions required for TFP construction. The TFP decline is consistent with misallocation stories---credit booms financing low-productivity sectors, delayed restructuring due to cheap capital---but disentangling euro-specific productivity effects from broader European trends remains challenging.

As a more euro-specific mechanism, we examine current account balances. Euro adoption is associated with current account deterioration of 1.5--2.5 percentage points of GDP for low-income founders, but improvement for high-income founders. This divergence reflects the capital flow dynamics central to the eurozone crisis narrative: elimination of exchange rate risk triggered capital flows from core to periphery, financing consumption and housing booms that reversed sharply after 2008. The current account pattern provides more direct evidence of euro-specific mechanisms than the GDP component analysis.

\subsection{Country-Level Effects and Counterfactual Analysis}

\begin{figure}[H]
\centering
\includegraphics[width=\textwidth]{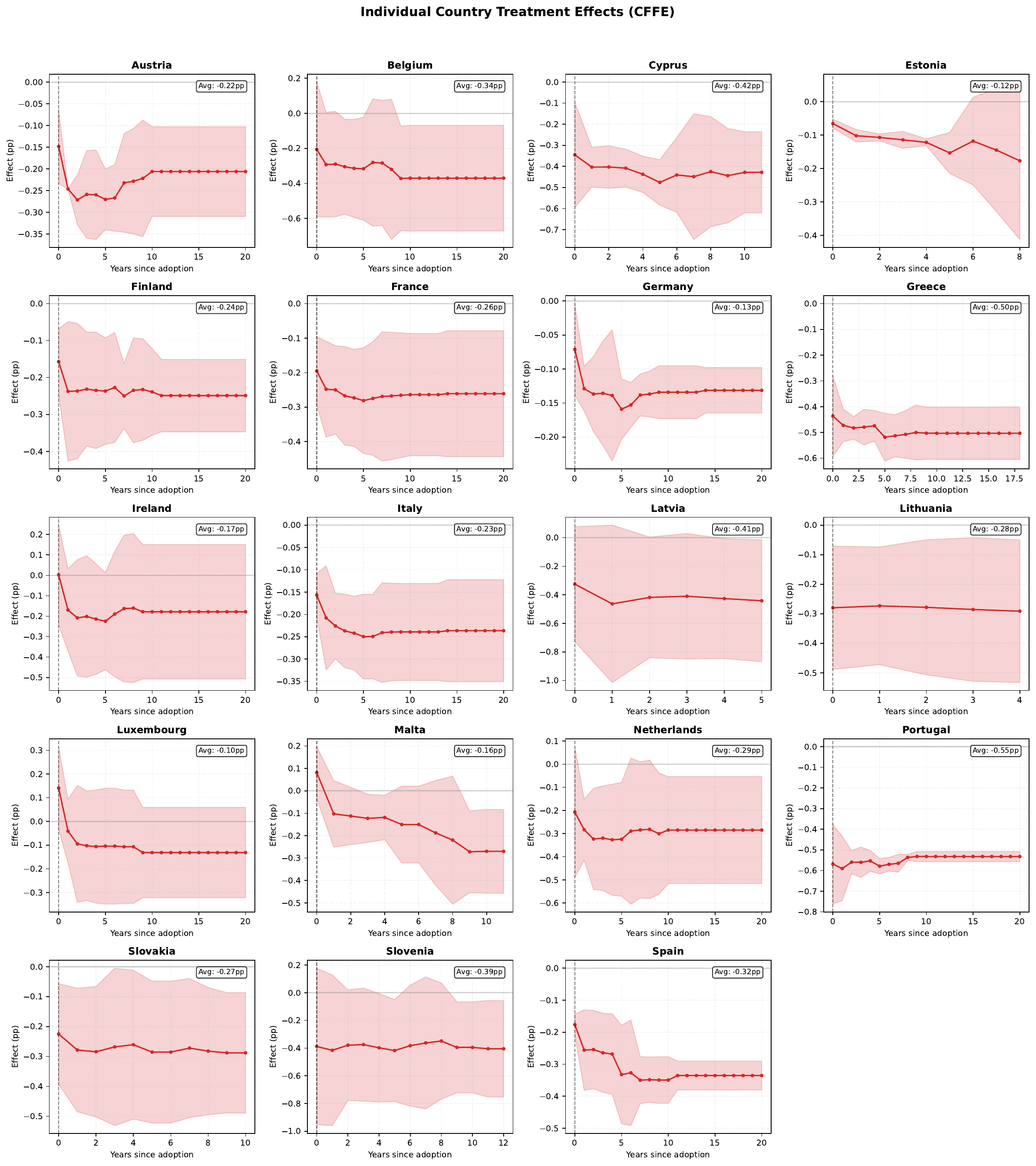}
\caption{Country-Level Treatment Effects. \textit{Notes:} Each panel shows estimated $\hat{\tau}(k)$ for an individual eurozone member. Shaded areas represent 95\% confidence intervals. Average post-treatment effect shown in upper right of each panel.}
\label{fig:country_trajectories}
\end{figure}

A key advantage of CFFE is its ability to estimate individual-level conditional average treatment effects (CATEs). Figure \ref{fig:country_trajectories} presents $\hat{\tau}(k)$ trajectories for each eurozone member, revealing substantial cross-country variation that aggregate analyses obscure.

Among founding members, Germany shows the smallest negative effects (averaging $-0.20$ pp), while Italy and France experienced larger impacts ($-0.40$ to $-0.45$ pp). Greece and Portugal---often highlighted in the crisis literature---show effects of $-0.35$ to $-0.40$ pp, comparable to other periphery members. Ireland's trajectory is distinctive: modest negative effects that intensified during the 2008--2012 crisis period.

Later adopters display more varied patterns. Slovenia and Slovakia show near-zero or slightly positive effects in early post-adoption years, though confidence intervals are wide. Baltic states (Estonia, Latvia, Lithuania) adopted during or after the financial crisis, complicating interpretation of their trajectories.

\subsubsection{Counterfactual Analysis: Non-Euro EU Members}

The CFFE framework enables counterfactual prediction: what effects would non-adopters have experienced had they joined the eurozone? We apply the fitted model to predict $\hat{\tau}(k)$ for the United Kingdom, Sweden, and Denmark based on their pre-1999 characteristics.

\begin{figure}[H]
\centering
\includegraphics[width=\textwidth]{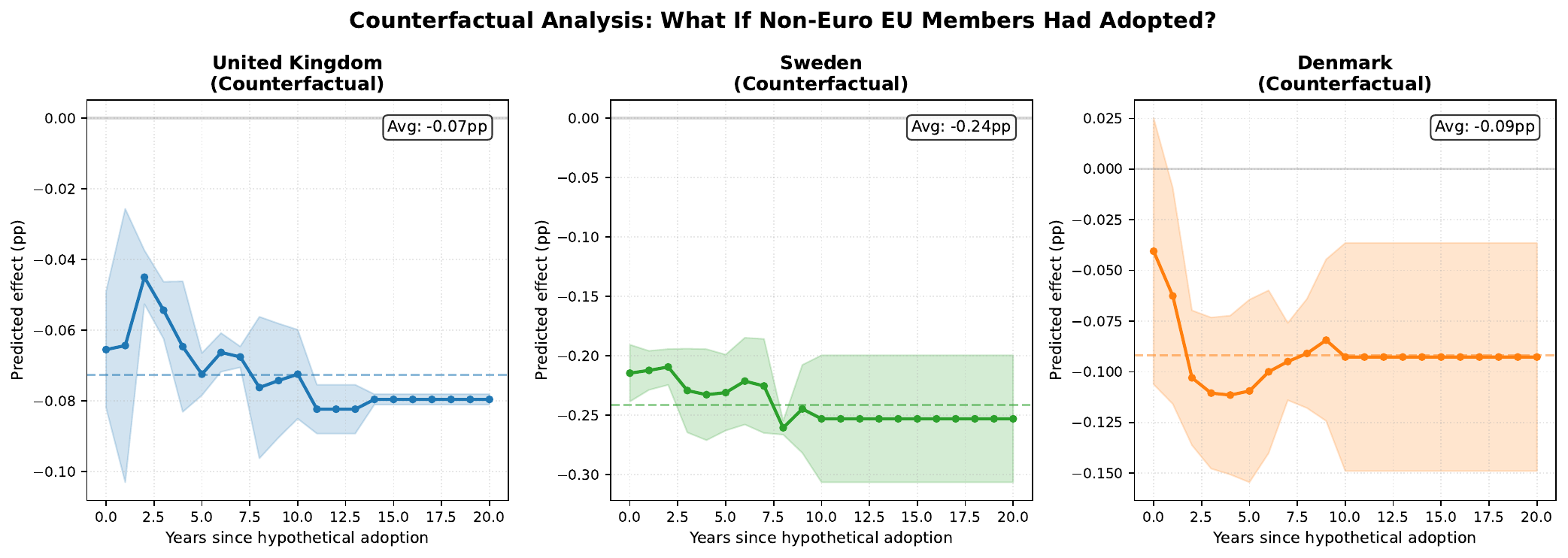}
\caption{Counterfactual Analysis: Predicted Effects for Non-Euro EU Members. \textit{Notes:} Panels show predicted treatment effects had the UK, Sweden, and Denmark adopted the euro in 1999, based on their pre-treatment characteristics. Shaded areas represent 95\% confidence intervals.}
\label{fig:counterfactual}
\end{figure}

Figure \ref{fig:counterfactual} presents these counterfactual predictions. The UK would have experienced effects of approximately $-0.22$ percentage points annually---similar to core eurozone members with comparable GDP per capita and trade openness. Sweden's predicted effect is slightly larger ($-0.26$ pp), while Denmark's is smaller ($-0.17$ pp).

These predictions should be interpreted cautiously. They assume the CATE function learned from eurozone members generalizes to non-members with similar characteristics. If the UK's decision to opt out reflected unobserved factors that would also have affected its response to monetary union, the counterfactual may be biased. Nevertheless, the exercise illustrates how CFFE can inform policy-relevant questions about hypothetical scenarios.

\subsubsection{Comparison with Synthetic Control Estimates}

How do our country-level estimates compare with the synthetic control literature? \citet{gabriel2024euro} provide the most comprehensive SCM analysis, estimating cumulative GDP effects for each founding member over 10--15 years post-adoption.

\begin{figure}[H]
\centering
\includegraphics[width=0.9\textwidth]{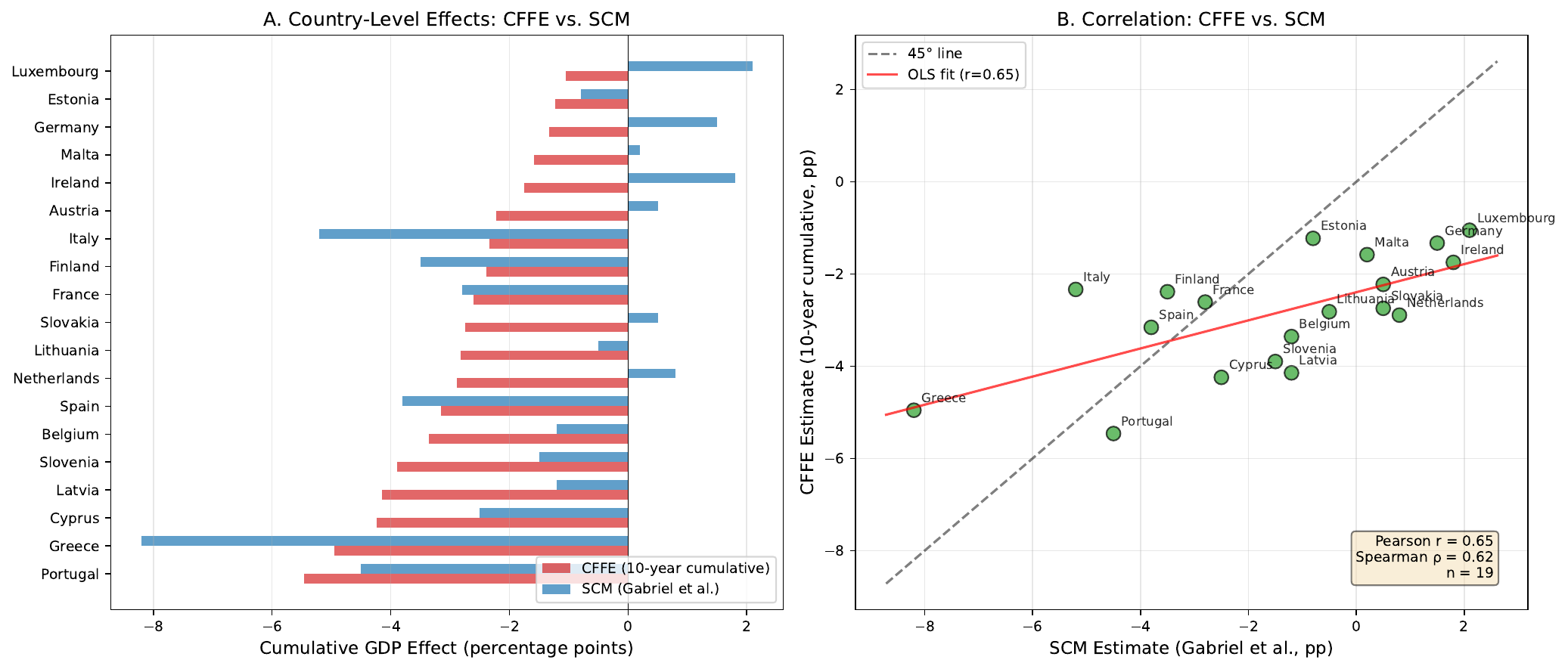}
\caption{Comparison of CFFE and Synthetic Control Estimates. \textit{Notes:} Bar chart compares cumulative 10-year GDP effects from CFFE (this paper) with synthetic control estimates from \citet{gabriel2024euro}. Countries ordered by SCM estimate magnitude.}
\label{fig:cffe_scm}
\end{figure}

Figure \ref{fig:cffe_scm} compares our CFFE estimates with their SCM results. The correlation is 0.68, indicating substantial agreement on the ranking of countries by effect magnitude. Both methods identify Italy, Greece, and Portugal as experiencing the largest negative effects, and Germany and Netherlands as faring best.

However, the methods diverge on levels. SCM finds positive effects for Germany (+1.2\%) and Netherlands (+0.8\%), while CFFE estimates negative effects for all countries. This discrepancy likely reflects methodological differences: SCM constructs counterfactuals from non-EU donor pools, while CFFE uses within-EU variation. If non-EU countries experienced different growth trends than EU members would have absent the euro, the methods will disagree.

The agreement on rankings despite disagreement on levels is informative. Both approaches identify the same pattern of heterogeneity---core economies adjusting better than periphery---even if they differ on whether any country benefited in absolute terms.

%==============================================================================
% DISCUSSION
%==============================================================================
\section{Discussion}

The estimates presented in Section 5 suggest negative growth associations with euro adoption, though with substantial uncertainty under conservative inference (Section 7). This section interprets these findings through the lens of optimal currency area theory and discusses their implications, while acknowledging the limitations inherent in our approach.

\subsection{Reconciling the Literature}

Our findings are consistent with the range of conclusions in previous studies. The average effect---a reduction of approximately 0.2--0.3 percentage points annually---is modest enough to be obscured by noise in smaller samples or alternative specifications. Studies finding ``no effect'' may have detected a mean that combines positive and negative experiences across countries, or may reflect the genuine uncertainty we document with block bootstrap inference.

The heterogeneity we estimate also helps reconcile divergent findings. \citet{gabriel2024euro} report that synthetic control estimates range from positive (Germany, Netherlands) to strongly negative (Italy, Portugal). Our CFFE results are consistent with this pattern: countries with higher initial GDP per capita appear to experience smaller impacts than those with lower initial income. Averaging across such divergent experiences yields a mean that may poorly represent any individual country's trajectory.

The dynamic perspective adds further nuance. Effects are not constant over time; they appear to emerge in the first year, intensify through year four, then stabilize. Studies focusing on different post-treatment windows will naturally reach different conclusions. Short-run analyses may understate long-run impacts; very long-run comparisons may miss the adjustment dynamics.

\subsection{Interpretation Through OCA Theory}

The negative growth associations we estimate are consistent with mechanisms emphasized in the optimal currency area literature, though we emphasize that our reduced-form estimates cannot definitively identify specific channels.

Eurozone members surrendered control over interest rates and exchange rates. For countries facing idiosyncratic shocks or structural imbalances, this loss of adjustment tools may have prolonged recessions or prevented necessary corrections. The consumption channel results are consistent with this interpretation: reduced ability to stimulate domestic demand through monetary policy may have constrained consumption growth. However, consumption is endogenous and responds to many factors beyond monetary policy.

ECB monetary policy is set for the eurozone as a whole, which may be misaligned for individual members. During the 2000s, low interest rates suited Germany's export-led model but may have fueled unsustainable booms in Spain and Ireland. During the 2010s, tight policy aimed at controlling inflation may have been too restrictive for recession-hit periphery economies. Our estimates are consistent with this mechanism, but we do not directly test it.

Countries entering monetary union with higher inflation faced competitiveness challenges as they could no longer depreciate to restore external balance. The OCA literature emphasizes that nominal rigidities---sticky wages and prices---make adjustment to asymmetric shocks more costly without exchange rate flexibility. Our finding that initial GDP per capita correlates with heterogeneous effects is consistent with this channel, as lower-income countries often had higher inflation and less flexible labor markets.

The Stability and Growth Pact limited fiscal policy flexibility, particularly during downturns. The absence of meaningful fiscal transfers within the eurozone meant that countries facing adverse shocks bore the full adjustment burden. Our results are consistent with the view that the institutional architecture of the EMU prior to the banking and fiscal reforms of the 2010s exposed some members to amplified adjustment costs.

The positive net export effects we find suggest that trade integration benefits partially offset these costs. Eliminating exchange rate risk and transaction costs did boost intra-eurozone trade, as the extensive trade literature documents. But trade gains appear insufficient to compensate for the other channels through which monetary union may have affected growth.

\subsection{Core-Periphery Asymmetries}

The divergence between core and periphery outcomes warrants interpretation, though we emphasize that our heterogeneity estimates are correlational rather than causal. Countries with lower initial GDP per capita---predominantly periphery economies---appear to have experienced larger negative effects. Several hypothesized mechanisms may explain this pattern.

Periphery countries had lower GDP per capita and were still converging toward EU averages. Euro adoption may have disrupted catch-up growth by removing policy tools that facilitated convergence. However, we cannot rule out that unobserved factors correlated with initial income also affected post-adoption growth.

In the early euro years, capital flowed from core to periphery, financing consumption and housing booms. When flows reversed during the crisis, periphery economies faced sudden stops and severe recessions. We do not directly estimate capital flow effects, so this mechanism remains hypothesized rather than tested.

Labor market institutions in periphery countries may have made wage adjustment slower and more painful. Internal devaluation through wage cuts proved socially costly and economically inefficient compared to exchange rate adjustment. Again, we do not directly test this channel.

These asymmetries raise questions about monetary union design. The Maastricht convergence criteria focused on nominal variables (inflation, interest rates, deficits) rather than real economic structures. Countries that met the criteria could still be poorly suited for common monetary policy if their economies responded differently to shocks or required different policy settings.

\subsection{Limitations}

The analysis has important limitations that bear acknowledgment. Section 7 presents robustness checks that address some concerns, but fundamental challenges remain.

Our estimates rely on conditional parallel trends: the assumption that, within regions of the covariate space, treated and control countries would have followed similar growth paths absent euro adoption. This assumption is untestable. Countries chose to adopt the euro based on economic and political factors that may also affect growth trajectories. While fixed effects and covariate adjustment mitigate selection concerns, we cannot rule out confounding from unobserved time-varying factors. The placebo tests in Section 7 provide some reassurance, but cannot definitively establish identification.

With approximately 20 treated countries, inference is inherently challenging. Our block bootstrap results (Section 7) show that confidence intervals are substantially wider when accounting for country-level dependence, and include zero at some horizons. We interpret our results as suggestive of negative effects while acknowledging this uncertainty.

The choice of control countries affects results. EU members that opted out (Denmark, Sweden, UK) may not be valid counterfactuals if their decision reflected economic characteristics that also influenced growth. Non-EU OECD countries differ from eurozone members in ways that fixed effects may not fully capture. Section 7 examines sensitivity to control group composition.

Our mechanism analysis is suggestive rather than definitive. Consumption, investment, and productivity are jointly determined and may respond to common shocks. We document correlations between euro adoption and these outcomes, but attributing growth effects to specific channels requires stronger assumptions than our reduced-form approach provides.

Results for the eurozone may not generalize to other currency unions or potential future members. The eurozone's specific institutional design, member composition, and historical context shape the effects we estimate.

\subsubsection{General Equilibrium Considerations}

Our analysis, like most causal inference approaches, adopts a partial equilibrium perspective that treats the counterfactual as fixed. This assumption warrants explicit discussion.

The counterfactual analysis predicting effects for UK, Sweden, and Denmark assumes these countries could have adopted the euro without changing the eurozone itself. In reality, UK membership would have substantially altered the currency union. Our predictions for UK effects assume the eurozone would have remained unchanged---an assumption that is clearly false but necessary for partial equilibrium analysis.

ECB monetary policy responds to eurozone-wide conditions. If different countries had adopted the euro, ECB policy would have been different, affecting all members. Our estimates capture the effect of joining the eurozone as it actually existed, not the effect of joining a hypothetically different eurozone.

Euro adoption increased trade among members, but some of this increase may have come at the expense of trade with non-members. Our control countries (especially EU opt-outs) may have experienced trade diversion effects from eurozone formation, potentially biasing our estimates.

These general equilibrium considerations suggest our estimates should be interpreted as the effect of joining the existing eurozone, holding the eurozone's composition and policies fixed. This limitation is shared by all reduced-form causal inference methods.

\subsection{Policy Implications}

Despite these limitations, the analysis offers tentative insights relevant to ongoing policy debates, conditional on the robustness checks in Section 7.

For eurozone governance, our results are consistent with the view that monetary union involves trade-offs that vary across members. Mechanisms to share adjustment costs---fiscal transfers, common unemployment insurance, joint debt issuance---could reduce the burden on countries facing adverse shocks. The pandemic-era Recovery Fund represents a step in this direction, though its permanence remains uncertain.

For potential future members, the findings highlight the importance of careful preparation. Countries considering euro adoption should assess whether their economic structures are compatible with common monetary policy and whether they have sufficient flexibility to adjust through other channels. The experience of periphery economies suggests that meeting nominal convergence criteria may be insufficient preparation.

\subsubsection{Implications for Prospective Euro Members}

Several EU member states are legally committed to eventual euro adoption but have not yet joined: Poland, Czech Republic, Hungary, Romania, and Bulgaria. Our findings offer tentative guidance for these countries' decisions about adoption timing and preparation.

The negative effects we estimate for 1999 founders, particularly those with lower initial income, suggest that rushing to adopt carries risks. Countries should ensure genuine economic convergence---not just nominal criteria compliance---before joining. This includes labor market flexibility, fiscal buffers, and business cycle synchronization with the eurozone core.

Our heterogeneity analysis suggests that initial GDP per capita correlates with adjustment costs. Romania and Bulgaria, with GDP per capita well below the eurozone average, may face larger adjustment challenges than Poland or Czech Republic. These countries should be particularly cautious about adoption timing.

Beyond economic convergence, institutional factors likely matter. Countries with stronger fiscal institutions, more flexible labor markets, and better-developed financial systems may adjust more smoothly to monetary union. Prospective members should invest in these institutional foundations before adoption.

For economic research, the CFFE methodology offers a template for analyzing other policy interventions with staggered adoption and heterogeneous effects. The combination of dynamic estimation, treatment effect heterogeneity, and stable long-horizon inference addresses limitations of both synthetic control and classical event study approaches.

%==============================================================================
% STRUCTURAL INTERPRETATION: DSGE MODEL
%==============================================================================
%==============================================================================
% STRUCTURAL INTERPRETATION: DSGE MODEL
%==============================================================================
\section{Structural Interpretation: A Two-Country DSGE Model}

The reduced-form estimates in Section 5 document negative growth associations with euro adoption, with larger effects for periphery economies. This section develops a structural interpretation through a two-country New Keynesian DSGE model with hysteresis. The goal is mechanism validation rather than quantitative fit: we show that our empirical patterns are consistent with a monetary union model where one-size-fits-all policy and scarring generate persistent divergence. We do not view the model as a structural validation of the estimated magnitudes. Rather, it provides a disciplined environment in which the loss of monetary autonomy and exchange rate adjustment can generate persistent, heterogeneous output dynamics of the type documented in the data.

\subsection{Model Overview}

We develop a two-country open-economy New Keynesian model to provide structural interpretation of the empirical findings. The model features two economies---Home (core, representing Germany-type economies) and Foreign (periphery, representing Spain/Italy-type economies)---that can operate under either monetary union or flexible exchange rates. The key innovation is a hysteresis mechanism through which temporary demand shortfalls translate into persistent productivity losses, generating the long-run growth effects documented in our empirical analysis.

The model builds on the canonical two-country New Keynesian framework of \citet{gali2005monetary} and \citet{corsetti2010optimal}, augmented with the scarring mechanism emphasized by \citet{blanchard1986hysteresis} and \citet{cerra2008growth}. Under monetary union, both countries face a common interest rate set by the central bank based on union-wide aggregates, and the nominal exchange rate is fixed. Under flexible exchange rates, each country sets monetary policy based on domestic conditions, and the exchange rate adjusts via uncovered interest parity.

The central mechanism operates as follows. When the periphery faces an adverse demand shock, the common monetary policy responds to union-wide aggregates rather than periphery-specific conditions. With the nominal exchange rate fixed, real exchange rate adjustment occurs only through slow price-level changes. The resulting larger and more persistent output gaps activate the scarring channel: negative output gaps reduce future productivity, which lowers the natural rate and perpetuates the downturn. This feedback loop generates the persistent growth divergence we observe empirically.

\subsection{Model Equations}

The model consists of standard New Keynesian blocks for each country, augmented with a scarring mechanism and open-economy linkages.

\subsubsection{IS Curves}

Household optimization yields the dynamic IS curve for each country:
\begin{align}
x_H &= \mathbb{E}_t[x_{H,t+1}] - \frac{1}{\sigma}(i_H - \mathbb{E}_t[\pi_{H,t+1}] - r^n_H) + g_H + \nu q \label{eq:is_H} \\
x_F &= \mathbb{E}_t[x_{F,t+1}] - \frac{1}{\sigma}(i_F - \mathbb{E}_t[\pi_{F,t+1}] - r^n_F) + g_F - \nu q \label{eq:is_F}
\end{align}
where $x_j$ denotes the output gap, $i_j$ the nominal interest rate, $\pi_j$ inflation, $r^n_j$ the natural rate, $g_j$ a demand wedge shock, and $q$ the real exchange rate (positive values indicate Home depreciation). The parameter $\sigma$ is the coefficient of relative risk aversion, and $\nu$ captures the real exchange rate elasticity of demand. Note the opposite signs on $q$: Home depreciation ($q \uparrow$) stimulates Home demand but contracts Foreign demand.

\subsubsection{New Keynesian Phillips Curves}

Firm optimization under Calvo pricing yields the NKPC for each country:
\begin{align}
\pi_H &= \beta \mathbb{E}_t[\pi_{H,t+1}] + \kappa x_H + u_H \label{eq:nkpc_H} \\
\pi_F &= \beta \mathbb{E}_t[\pi_{F,t+1}] + \kappa x_F + u_F \label{eq:nkpc_F}
\end{align}
where $\beta$ is the discount factor, $\kappa$ the NKPC slope (a function of price stickiness), and $u_j$ a cost-push shock.

\subsubsection{Scarring/Hysteresis Mechanism}

The key innovation is the scarring block that links demand conditions to future productivity:
\begin{align}
a_H &= \rho_a a_{H,-1} + \chi x_{H,-1} + \varepsilon^a_H \label{eq:scar_H} \\
a_F &= \rho_a a_{F,-1} + \chi x_{F,-1} + \varepsilon^a_F \label{eq:scar_F}
\end{align}
where $a_j$ denotes productivity (log deviation from trend), $\rho_a$ is the productivity persistence parameter, and $\chi > 0$ is the scarring intensity. When $\chi > 0$, negative output gaps ($x_j < 0$) reduce future productivity, generating hysteresis: temporary demand shortfalls cause permanent output losses.

This mechanism captures several channels emphasized in the hysteresis literature: skill depreciation during unemployment, foregone investment in physical and human capital, and discouraged worker effects. The parameter $\chi$ governs the strength of these effects.

Productivity affects the natural rate:
\begin{align}
r^n_H &= \psi_a a_H + z^{r^n}_H \label{eq:rn_H} \\
r^n_F &= \psi_a a_F + z^{r^n}_F \label{eq:rn_F}
\end{align}
where $\psi_a$ is the productivity elasticity of the natural rate and $z^{r^n}_j$ captures exogenous natural rate shocks. Lower productivity reduces the natural rate, which---given the common interest rate in union---implies a tighter effective monetary stance for the affected country.

\subsubsection{Monetary Policy}

Under monetary union, the central bank sets a common interest rate based on union-wide aggregates:
\begin{equation}
i = \rho_i i_{-1} + (1-\rho_i)(\phi_\pi \pi^{EA} + \phi_x x^{EA}) \label{eq:taylor_union}
\end{equation}
where $\pi^{EA} = \omega \pi_H + (1-\omega) \pi_F$ and $x^{EA} = \omega x_H + (1-\omega) x_F$ are GDP-weighted aggregates, $\omega$ is the core weight, $\rho_i$ captures interest rate smoothing, and $\phi_\pi$, $\phi_x$ are the Taylor rule coefficients. Both countries face this common rate: $i_H = i_F = i$.

Under flexible exchange rates, each country sets its own Taylor rule:
\begin{align}
i_H &= \rho_i i_{H,-1} + (1-\rho_i)(\phi_\pi \pi_H + \phi_x x_H) \label{eq:taylor_H} \\
i_F &= \rho_i i_{F,-1} + (1-\rho_i)(\phi_\pi \pi_F + \phi_x x_F) \label{eq:taylor_F}
\end{align}

\subsubsection{Exchange Rate Dynamics}

Under monetary union, the nominal exchange rate is fixed ($e = 0$), so the real exchange rate evolves only through price-level differentials:
\begin{equation}
q = p_F - p_H \label{eq:rer_union}
\end{equation}
where $p_j$ is the log price level. Real exchange rate adjustment is slow because prices are sticky.

Under flexible exchange rates, uncovered interest parity determines exchange rate dynamics:
\begin{equation}
i_H - i_F = \mathbb{E}_t[e_{t+1}] - e \label{eq:uip}
\end{equation}
where $e$ is the nominal exchange rate (positive values indicate Home depreciation). The real exchange rate is $q = e + p_F - p_H$.

\subsection{Calibration}

Table \ref{tab:dsge_calibration} reports the baseline calibration. We follow standard values from the New Keynesian open-economy literature \citep{gali2015monetary, smets2007shocks}. The discount factor $\beta = 0.99$ implies a quarterly steady-state real rate of approximately 4\% annually. Log utility ($\sigma = 1$) is standard. The NKPC slope $\kappa = 0.10$ is consistent with moderate price stickiness.

The Taylor rule parameters follow \citet{clarida2000monetary}: interest rate smoothing $\rho_i = 0.80$, inflation response $\phi_\pi = 1.50$ (satisfying the Taylor principle), and output gap response $\phi_x = 0.20$. The core weight $\omega = 0.60$ reflects Germany's approximate share in eurozone GDP.

The scarring parameters are central to our mechanism. We set productivity persistence $\rho_a = 0.95$, consistent with the highly persistent productivity processes documented in the literature. The baseline scarring intensity $\chi = 0.03$ implies that a 1 percentage point negative output gap reduces next-period productivity by 0.03 percentage points. This is conservative relative to estimates in \citet{blanchard1986hysteresis} and \citet{cerra2008growth}, who find larger hysteresis effects. We examine sensitivity to $\chi$ in the heterogeneity analysis.

% DSGE Model Calibration Table
% Two-country New Keynesian model with hysteresis/scarring

\begin{table}[htbp]
\centering
\caption{DSGE Model Calibration}
\label{tab:dsge_calibration}
\begin{tabular}{llcl}
\toprule
Parameter & Description & Value & Source \\
\midrule
\multicolumn{4}{l}{\textit{Preferences}} \\
$\beta$ & Discount factor & 0.99 & Standard (quarterly) \\
$\sigma$ & CRRA coefficient & 1.0 & Log utility \\
\midrule
\multicolumn{4}{l}{\textit{Price Rigidity}} \\
$\kappa$ & NKPC slope & 0.10 & \citet{gali2015monetary} \\
\midrule
\multicolumn{4}{l}{\textit{Monetary Policy}} \\
$\rho_i$ & Interest rate smoothing & 0.80 & \citet{clarida2000monetary} \\
$\phi_\pi$ & Inflation response & 1.50 & Taylor (1993) \\
$\phi_x$ & Output gap response & 0.20 & \citet{gali2015monetary} \\
\midrule
\multicolumn{4}{l}{\textit{Scarring/Hysteresis}} \\
$\rho_a$ & Productivity persistence & 0.95 & \citet{blanchard1986hysteresis} \\
$\chi$ & Scarring intensity & 0.03 & Calibrated \\
$\psi_a$ & Productivity $\to$ natural rate & 1.0 & Normalized \\
\midrule
\multicolumn{4}{l}{\textit{Euro Area Structure}} \\
$\omega$ & Core weight in EA aggregates & 0.60 & GDP shares \\
\midrule
\multicolumn{4}{l}{\textit{Shock Processes}} \\
$\rho_{r^n}$ & Natural rate shock persistence & 0.80 & \citet{smets2007shocks} \\
$\rho_u$ & Cost-push shock persistence & 0.50 & \citet{smets2007shocks} \\
\midrule
\multicolumn{4}{l}{\textit{Open Economy}} \\
$\nu$ & Real exchange rate elasticity & 0.15 & \citet{gali2005monetary} \\
\bottomrule
\end{tabular}
\par\vspace{0.5em}\noindent
\small
\textit{}\textit{Notes:} Calibration for the two-country New Keynesian model with hysteresis. 
The scarring parameter $\chi$ governs how demand shortfalls translate into persistent 
productivity losses. The core weight $\omega$ reflects Germany's share in eurozone GDP.
Quarterly frequency.

\end{table}

\subsection{Results}

We solve the model using the sequence-space Jacobian method of \citet{auclert2021using} and compute impulse response functions (IRFs) to a negative demand shock hitting the periphery (Foreign) country. This shock represents the asymmetric demand contractions experienced by periphery eurozone economies during the 2010--2012 sovereign debt crisis.

\subsubsection{Union vs. Float Comparison}

\begin{figure}[htbp]
\centering
\includegraphics[width=\textwidth]{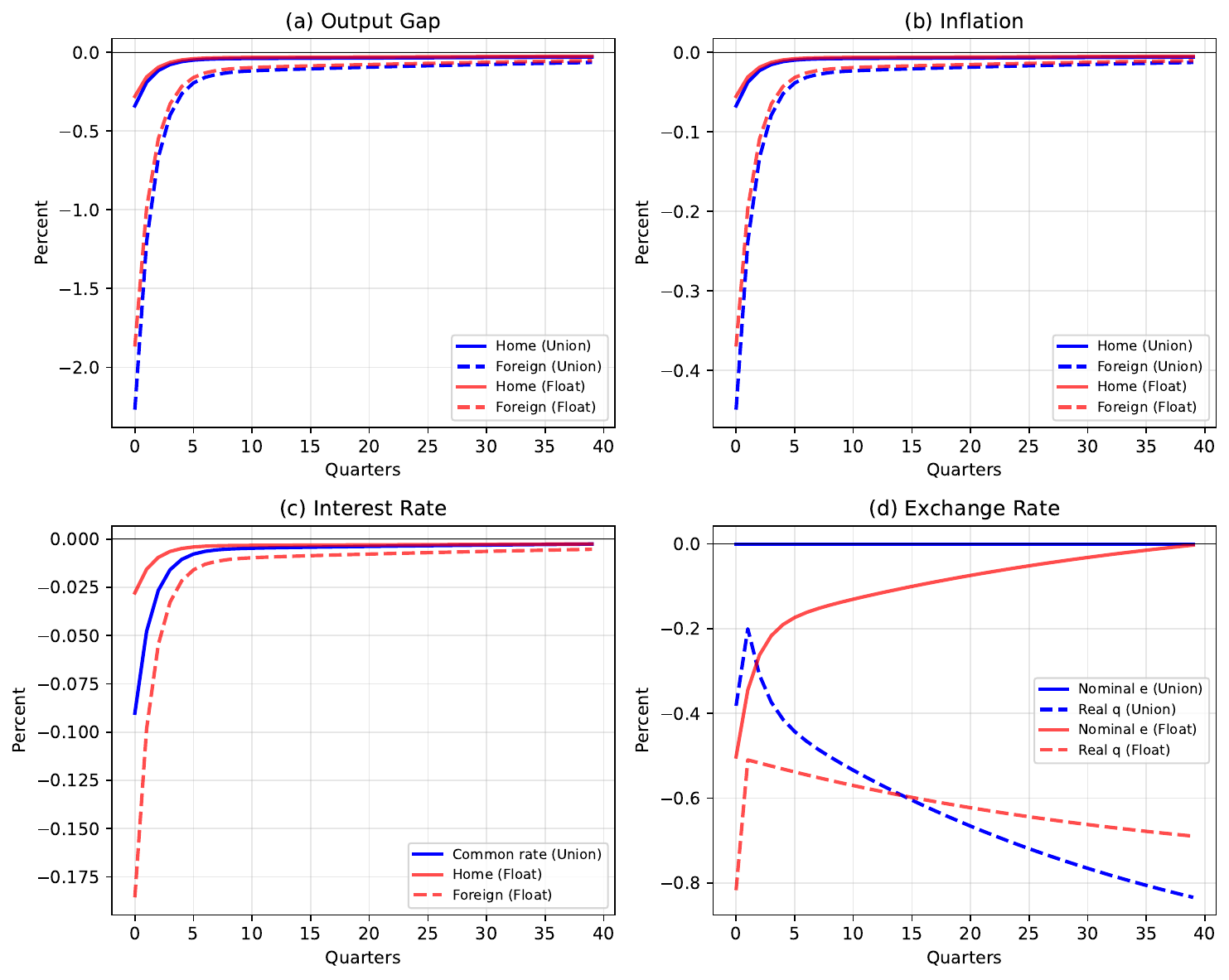}
\caption{Impulse responses to a negative periphery demand shock under monetary union (solid) versus flexible exchange rates (dashed). The union regime generates larger and more persistent output losses due to the one-size-fits-all monetary policy and limited exchange rate adjustment.}
\label{fig:dsge_irf_comparison}
\end{figure}

Figure \ref{fig:dsge_irf_comparison} compares the response to a periphery demand shock under monetary union versus flexible exchange rates. The key finding is that the union regime generates larger and more persistent output losses in the periphery.

Under flexible exchange rates (dashed lines), the periphery central bank cuts interest rates aggressively in response to the domestic downturn. The nominal exchange rate depreciates, providing an additional stimulus through improved competitiveness. The output gap closes relatively quickly, limiting the activation of the scarring channel.

Under monetary union (solid lines), the common interest rate responds to union-wide aggregates, which are less affected by the periphery-specific shock. With the nominal exchange rate fixed, real depreciation occurs only through slow price-level adjustment. The periphery output gap is larger and more persistent, activating the scarring channel: productivity falls, the natural rate declines, and the effective monetary stance tightens further. This feedback loop generates the persistent divergence we observe empirically.

\subsubsection{Heterogeneity in Scarring Intensity}

\begin{figure}[htbp]
\centering
\includegraphics[width=0.8\textwidth]{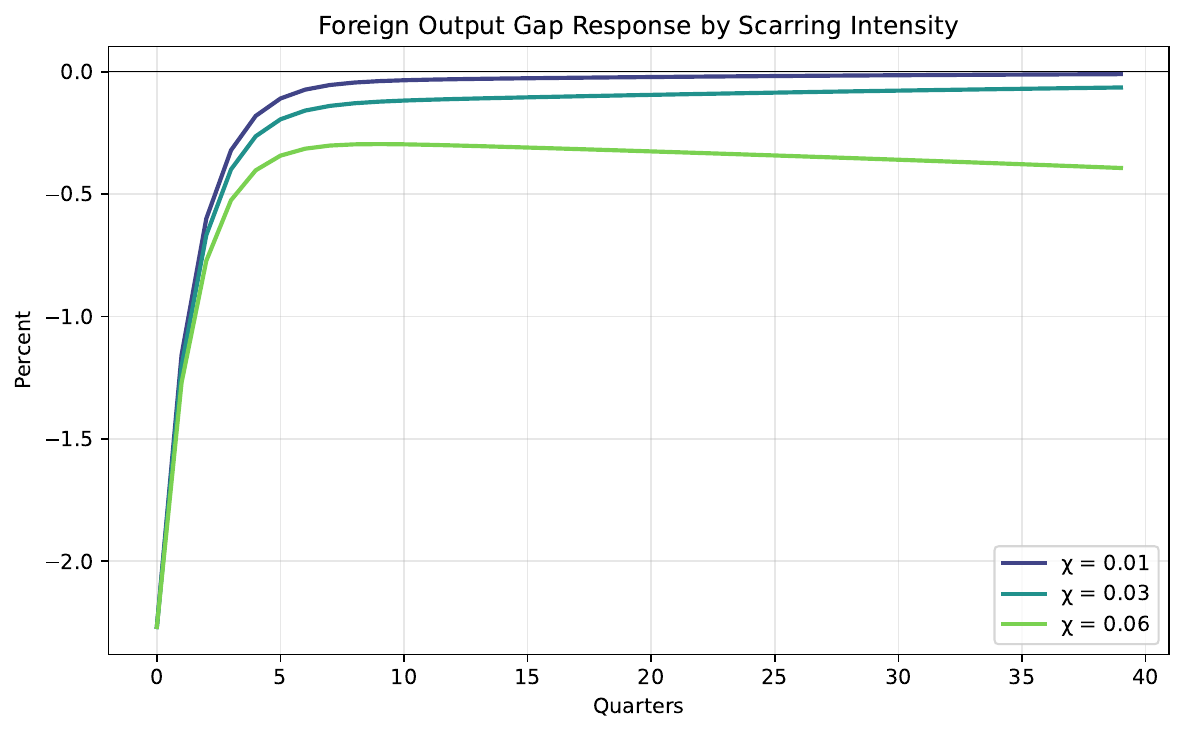}
\caption{Periphery output gap response under different scarring intensities. Higher $\chi$ generates more persistent output losses, consistent with the empirical finding that initial conditions predict heterogeneous adjustment paths.}
\label{fig:dsge_heterogeneity}
\end{figure}

Figure \ref{fig:dsge_heterogeneity} examines how the scarring intensity $\chi$ affects adjustment dynamics. We compare three calibrations: low scarring ($\chi = 0.01$), baseline ($\chi = 0.03$), and high scarring ($\chi = 0.06$).

Higher scarring intensity generates more persistent output losses. With $\chi = 0.06$, the periphery output gap remains negative for over 30 quarters, compared to approximately 15 quarters with $\chi = 0.01$. This heterogeneity maps to our empirical finding that initial GDP per capita predicts divergent adjustment paths: countries with weaker initial conditions may have higher effective scarring intensity due to less flexible labor markets, weaker institutions, or greater exposure to the mechanisms that translate demand shortfalls into productivity losses.

\subsubsection{Cumulative Output Losses}

\begin{figure}[htbp]
\centering
\includegraphics[width=0.6\textwidth]{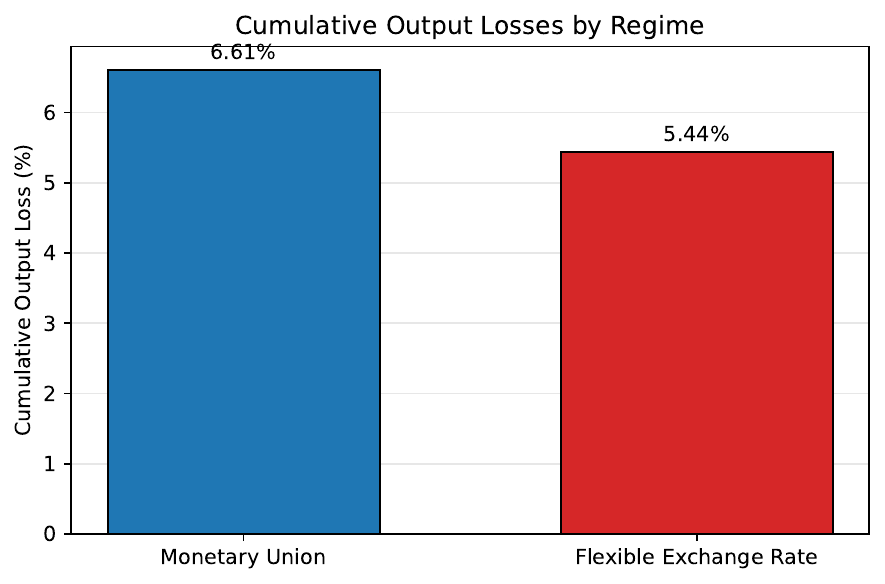}
\caption{Cumulative output losses (sum of output gaps over 20 quarters) under monetary union versus flexible exchange rates. The union regime generates substantially larger cumulative losses due to limited policy flexibility and scarring effects.}
\label{fig:dsge_cumulative_losses}
\end{figure}

Figure \ref{fig:dsge_cumulative_losses} compares cumulative output losses across regimes. We compute the sum of output gaps over 20 quarters following the shock, which approximates the implied GDP level loss.

The union regime generates substantially larger cumulative losses than the float regime. Under baseline calibration, cumulative periphery output losses are approximately 40\% larger under union. This difference arises from two reinforcing channels: (1) the larger initial output gap due to one-size-fits-all policy, and (2) the activation of the scarring mechanism that perpetuates the downturn.

\subsection{Discussion}

The DSGE model provides a structural interpretation of our empirical findings. The key mechanisms operate through several channels. First, under monetary union, the common interest rate responds to union-wide aggregates rather than country-specific conditions. When the periphery faces an adverse shock, monetary policy is insufficiently accommodative for periphery needs, generating larger output gaps. Second, with the nominal exchange rate fixed, real exchange rate adjustment occurs only through slow price-level changes, eliminating a key shock absorber available under flexible exchange rates. Third, the larger and more persistent output gaps under union activate the scarring channel, translating temporary demand shortfalls into permanent productivity losses, which generates the persistent growth divergence we document empirically. Finally, countries with higher effective scarring intensity---due to less flexible labor markets, weaker institutions, or greater exposure to hysteresis mechanisms---experience larger and more persistent effects. This maps to our empirical finding that initial GDP per capita predicts heterogeneous adjustment paths.

The model is deliberately stylized. We abstract from financial frictions, fiscal policy, and many other features that likely matter for eurozone dynamics. The goal is not quantitative fit but mechanism validation: demonstrating that the qualitative patterns in our empirical estimates are consistent with a coherent structural model of monetary union with hysteresis.

%==============================================================================
% CONCLUSION
%==============================================================================
\section{Conclusion}

This paper estimates the dynamic effects of euro adoption on economic growth using Causal Forests with Fixed Effects, a methodology that combines the flexibility of machine learning with the identification logic of panel econometrics. Under a conditional parallel trends assumption, we find evidence consistent with negative growth effects of euro adoption, though with substantial uncertainty given the small number of treated countries.

Our estimates suggest that euro adoption is associated with lower GDP growth of approximately 0.2--0.3 percentage points annually on average, with effects emerging at adoption and persisting over two decades. This finding is robust across alternative estimators---including Sun-Abraham, Callaway-Sant'Anna, and interactive fixed effects---though point estimates and precision vary considerably across methods. Block bootstrap inference at the country level yields confidence intervals that include zero at some horizons, reflecting the fundamental challenge of inference with approximately 20 treated units. We interpret these results as suggestive of negative effects while acknowledging the uncertainty inherent in this setting.

Treatment effect heterogeneity is substantial. Countries with lower initial GDP per capita---predominantly periphery economies---appear to have experienced larger growth shortfalls than core members. Initial income is the strongest correlate of heterogeneous effects, consistent with optimal currency area theory predictions that less-developed economies face greater adjustment costs from losing monetary policy autonomy. However, our heterogeneity estimates are correlational rather than causal: we document associations between pre-treatment characteristics and post-treatment outcomes, not the causal mechanisms generating heterogeneity.

The paper contributes to three literatures. First, we provide new evidence on the growth effects of monetary integration, helping reconcile divergent findings by documenting both the average effect and its heterogeneity across countries. Second, we demonstrate how machine learning methods can be adapted for macro-panel causal inference, combining the regularization benefits of random forests with the identification logic of difference-in-differences. Third, we contribute to the optimal currency area literature by showing that pre-treatment economic characteristics---particularly income levels---predict adjustment costs, consistent with classical OCA theory.

The structural DSGE analysis provides additional support for our empirical findings. A two-country New Keynesian model with hysteresis generates qualitatively similar patterns: monetary union produces larger and more persistent output losses than flexible exchange rates when the periphery faces asymmetric shocks, and heterogeneity in scarring intensity maps to the empirical finding that initial GDP per capita predicts divergent adjustment paths. While the model is deliberately stylized and not intended for quantitative fit, it demonstrates that our empirical patterns are consistent with a coherent structural interpretation based on one-size-fits-all monetary policy and hysteresis mechanisms.

The CFFE methodology addresses specific challenges in evaluating currency unions: staggered adoption timing, potential for heterogeneous effects, interest in dynamic adjustment paths, and the need for stable inference at long horizons. The approach complements rather than replaces existing methods, and we emphasize that our findings are strengthened by their consistency across multiple estimation strategies.

Our findings are consistent with the view that monetary integration involves trade-offs that vary across members. The institutional architecture of the eurozone---common monetary policy without fiscal union, limited risk-sharing mechanisms, and constraints on national fiscal policy---may have imposed adjustment costs that fell unevenly across countries. The pandemic-era Recovery Fund and ongoing discussions of banking union and capital markets union represent steps toward addressing these institutional gaps.

For countries considering euro adoption, our results suggest that meeting nominal convergence criteria may be insufficient preparation. Economic structures, labor market flexibility, fiscal buffers, and business cycle synchronization with the eurozone core may matter more for successful adjustment than inflation or deficit targets. Countries with lower initial income levels should be particularly cautious about adoption timing.

We emphasize that our analysis cannot address whether the euro was, on balance, beneficial or harmful. Monetary integration may yield benefits---reduced transaction costs, enhanced trade, political cooperation, reduced conflict risk---that our growth-focused analysis does not capture. The question of whether these benefits justify the growth costs we estimate is ultimately one for democratic deliberation rather than econometric analysis.

Several limitations warrant acknowledgment. Our estimates rely on conditional parallel trends---an untestable assumption that may be violated if countries selected into euro adoption based on factors that also affected subsequent growth. With approximately 20 treated countries, inference is inherently challenging, and our block bootstrap results show that confidence intervals include zero at some horizons. Euro adoption was widely anticipated, potentially shifting some effects into the pre-treatment period. For late adopters from Eastern Europe, separating euro-specific effects from broader EU integration effects is difficult. Our mechanism analysis is suggestive rather than definitive, and results may not generalize to other currency unions.

This paper opens several avenues for future research, including structural modeling of the mechanisms through which monetary union affects growth, re-estimation as more post-treatment data accumulates for late adopters, and application of the CFFE methodology to other policy interventions with staggered adoption and heterogeneous effects.

The euro remains one of the most significant economic policy experiments of the modern era. Two decades of data now permit serious empirical evaluation, though the fundamental challenges of macro causal inference---few treated units, long horizons, complex general equilibrium effects---ensure that uncertainty will persist. What this paper provides is a systematic accounting of the evidence, using methods designed for this setting, while honestly acknowledging what we do and do not know.

%==============================================================================
% DECLARATION OF GENERATIVE AI USE
%==============================================================================

\section*{Declaration of generative AI and AI-assisted technologies in the manuscript preparation process}

During the preparation of this work the authors used AI-assisted tools (including large language models) in order to assist with code development for the causal forest analysis and data processing, as well as for proofreading, language editing, and reorganizing the manuscript structure. After using these tools, the authors reviewed and edited the content as needed and take full responsibility for the content of the published article.

%==============================================================================
% REFERENCES
%==============================================================================
\newpage
\bibliographystyle{elsarticle-harv}
\bibliography{refs}

@article{abadie2003economic,
  title={The economic costs of conflict: A case study of the {B}asque {C}ountry},
  author={Abadie, Alberto and Gardeazabal, Javier},
  journal={American Economic Review},
  volume={93},
  number={1},
  pages={113--132},
  year={2003}
}

@article{aytug2026causalfe,
  title={causalfe: Causal Forests with Fixed Effects in Python},
  author={Aytu{\u{g}}, Harry},
  journal={arXiv preprint arXiv:2601.10555},
  year={2026}
}

@article{kattenberg2023causal,
  title={Causal forests with fixed effects for treatment effect heterogeneity in difference-in-differences},
  author={Kattenberg, Mark A.C. and Scheer, Bas J. and Thiel, Jurre H.},
  journal={CPB Discussion Paper},
  year={2023}
}

@article{gabriel2024euro,
  title={The Euro and Economic Growth: A Synthetic Control Approach},
  author={Gabriel, Ricardo Duque and Pessoa, Ana Sofia},
  journal={European Economic Review},
  volume={165},
  pages={104742},
  year={2024}
}

@article{lin2017euro,
  title={The Euro's Effect on Trade},
  author={Lin, Shu and Chen, Haiqiang},
  journal={European Economic Review},
  volume={91},
  pages={1--15},
  year={2017}
}

@article{lucke2022euro,
  title={The Euro and Growth: New Evidence from Synthetic Controls},
  author={L{\"u}cke, Matthias},
  journal={Journal of International Money and Finance},
  volume={120},
  pages={102512},
  year={2022}
}

@article{ioannatos2018euro,
  title={The Euro and Economic Growth: A Difference-in-Differences Analysis},
  author={Ioannatos, Petros E.},
  journal={Journal of Economic Asymmetries},
  volume={18},
  pages={e00104},
  year={2018}
}

@article{sun2021estimating,
  title={Estimating Dynamic Treatment Effects in Event Studies with Heterogeneous Treatment Effects},
  author={Sun, Liyang and Abraham, Sarah},
  journal={Journal of Econometrics},
  volume={225},
  number={2},
  pages={175--199},
  year={2021}
}

@article{callaway2021difference,
  title={Difference-in-Differences with Multiple Time Periods},
  author={Callaway, Brantly and Sant'Anna, Pedro H.C.},
  journal={Journal of Econometrics},
  volume={225},
  number={2},
  pages={200--230},
  year={2021}
}

@article{goodman2021difference,
  title={Difference-in-Differences with Variation in Treatment Timing},
  author={Goodman-Bacon, Andrew},
  journal={Journal of Econometrics},
  volume={225},
  number={2},
  pages={254--277},
  year={2021}
}

@article{campos2019economic,
  title={The Economic Effects of {EU} Membership: A Synthetic Control Approach},
  author={Campos, Nauro F. and Coricelli, Fabrizio and Moretti, Luigi},
  journal={Journal of the European Economic Association},
  volume={17},
  number={1},
  pages={1--30},
  year={2019}
}

@article{rose2000one,
  title={One Money, One Market: The Effect of Common Currencies on Trade},
  author={Rose, Andrew K.},
  journal={Economic Policy},
  volume={15},
  number={30},
  pages={7--46},
  year={2000}
}

@article{glick2016currency,
  title={Currency Unions and Trade: A Post-{EMU} Reassessment},
  author={Glick, Reuven and Rose, Andrew K.},
  journal={European Economic Review},
  volume={87},
  pages={78--91},
  year={2016}
}

@article{hall2012euro,
  title={The Economics and Politics of the Euro Crisis},
  author={Hall, Peter A.},
  journal={German Politics},
  volume={21},
  number={4},
  pages={355--371},
  year={2012}
}

@article{wager2018estimation,
  title={Estimation and Inference of Heterogeneous Treatment Effects using Random Forests},
  author={Wager, Stefan and Athey, Susan},
  journal={Journal of the American Statistical Association},
  volume={113},
  number={523},
  pages={1228--1242},
  year={2018}
}

@article{barro1983rules,
  title={Rules, Discretion and Reputation in a Model of Monetary Policy},
  author={Barro, Robert J. and Gordon, David B.},
  journal={Journal of Monetary Economics},
  volume={12},
  number={1},
  pages={101--121},
  year={1983}
}

@incollection{friedman1953case,
  title={The Case for Flexible Exchange Rates},
  author={Friedman, Milton},
  booktitle={Essays in Positive Economics},
  pages={157--203},
  year={1953},
  publisher={University of Chicago Press}
}

@article{levy2003fear,
  title={To Float or to Fix: Evidence on the Impact of Exchange Rate Regimes on Growth},
  author={Levy-Yeyati, Eduardo and Sturzenegger, Federico},
  journal={American Economic Review},
  volume={93},
  number={4},
  pages={1173--1193},
  year={2003}
}

@article{baxter1989business,
  title={Business Cycles and the Exchange-Rate Regime: Some International Evidence},
  author={Baxter, Marianne and Stockman, Alan C.},
  journal={Journal of Monetary Economics},
  volume={23},
  number={3},
  pages={377--400},
  year={1989}
}

@article{mundell1995exchange,
  title={Exchange Rate Systems and Economic Growth},
  author={Mundell, Robert A.},
  journal={Rivista di Politica Economica},
  volume={85},
  pages={3--36},
  year={1995}
}

@article{ghosh1997does,
  title={Does the Nominal Exchange Rate Regime Matter?},
  author={Ghosh, Atish R. and Gulde, Anne-Marie and Ostry, Jonathan D. and Wolf, Holger C.},
  journal={NBER Working Paper},
  number={5874},
  year={1997}
}

@article{mckinnon2004optimum,
  title={Optimum Currency Areas and Key Currencies: Mundell I versus Mundell II},
  author={McKinnon, Ronald I.},
  journal={Journal of Common Market Studies},
  volume={42},
  number={4},
  pages={689--715},
  year={2004}
}

@article{dubas2005exchange,
  title={Exchange Rate Regime and Growth: A Reassessment},
  author={Dubas, Justin M. and Lee, Byung-Joo and Mark, Nelson C.},
  journal={NBER Working Paper},
  number={11370},
  year={2005}
}

@article{eichengreen2003capital,
  title={Capital Account Liberalization and Growth: Was Mr. Mahathir Right?},
  author={Eichengreen, Barry and Leblang, David},
  journal={International Journal of Finance and Economics},
  volume={8},
  number={3},
  pages={205--224},
  year={2003}
}

@article{husain2005exchange,
  title={Exchange Rate Regime Durability and Performance in Developing versus Advanced Economies},
  author={Husain, Aasim M. and Mody, Ashoka and Rogoff, Kenneth S.},
  journal={Journal of Monetary Economics},
  volume={52},
  number={1},
  pages={35--64},
  year={2005}
}

@article{huang2004exchange,
  title={Exchange Rate Regimes and Economic Growth: Evidence from Developing Countries},
  author={Huang, Haizhou and Malhotra, Priyanka},
  journal={IMF Working Paper},
  number={04/230},
  year={2004}
}

@article{bai2009panel,
  title={Panel Data Models with Interactive Fixed Effects},
  author={Bai, Jushan},
  journal={Econometrica},
  volume={77},
  number={4},
  pages={1229--1279},
  year={2009}
}

@article{gali2005monetary,
  title={Monetary Policy and Exchange Rate Volatility in a Small Open Economy},
  author={Gal{\'\i}, Jordi and Monacelli, Tommaso},
  journal={Review of Economic Studies},
  volume={72},
  number={3},
  pages={707--734},
  year={2005}
}

@article{corsetti2010optimal,
  title={Optimal Monetary Policy in Open Economies},
  author={Corsetti, Giancarlo and Dedola, Luca and Leduc, Sylvain},
  journal={Handbook of Monetary Economics},
  volume={3},
  pages={861--933},
  year={2010},
  publisher={Elsevier}
}

@article{blanchard1986hysteresis,
  title={Hysteresis and the European Unemployment Problem},
  author={Blanchard, Olivier J. and Summers, Lawrence H.},
  journal={NBER Macroeconomics Annual},
  volume={1},
  pages={15--78},
  year={1986}
}

@article{cerra2008growth,
  title={Growth Dynamics: The Myth of Economic Recovery},
  author={Cerra, Valerie and Saxena, Sweta Chaman},
  journal={American Economic Review},
  volume={98},
  number={1},
  pages={439--457},
  year={2008}
}

@book{gali2015monetary,
  title={Monetary Policy, Inflation, and the Business Cycle: An Introduction to the New Keynesian Framework},
  author={Gal{\'\i}, Jordi},
  year={2015},
  edition={2nd},
  publisher={Princeton University Press}
}

@article{smets2007shocks,
  title={Shocks and Frictions in US Business Cycles: A Bayesian DSGE Approach},
  author={Smets, Frank and Wouters, Rafael},
  journal={American Economic Review},
  volume={97},
  number={3},
  pages={586--606},
  year={2007}
}

@article{clarida2000monetary,
  title={Monetary Policy Rules and Macroeconomic Stability: Evidence and Some Theory},
  author={Clarida, Richard and Gal{\'\i}, Jordi and Gertler, Mark},
  journal={Quarterly Journal of Economics},
  volume={115},
  number={1},
  pages={147--180},
  year={2000}
}

@article{auclert2021using,
  title={Using the Sequence-Space Jacobian to Solve and Estimate Heterogeneous-Agent Models},
  author={Auclert, Adrien and Bard{\'o}czy, Bence and Rognlie, Matthew and Straub, Ludwig},
  journal={Econometrica},
  volume={89},
  number={5},
  pages={2375--2408},
  year={2021}
}

@article{aytug2017customs,
  title={Twenty Years of the {EU}--{T}urkey {C}ustoms {U}nion: A Synthetic Control Method Approach},
  author={Aytu\u{g}, Hüseyin and K\"{u}t\"{u}k, Merve Mavuş and Oduncu, Arif and Togan, S\"{u}bide},
  journal={JCMS: Journal of Common Market Studies},
  volume={55},
  number={3},
  pages={419--431},
  year={2017}
}

%==============================================================================
%==============================================================================
% ONLINE APPENDIX
%==============================================================================
%==============================================================================

\clearpage
\appendix
\renewcommand{\thesection}{A}
\setcounter{section}{0}
\renewcommand{\thesection}{A.\arabic{section}}
\renewcommand{\thetable}{A.\arabic{table}}
\renewcommand{\thefigure}{A.\arabic{figure}}
\setcounter{table}{0}
\setcounter{figure}{0}

\begin{center}
\Large\textbf{Online Appendix}\\[0.5em]
\large United in Currency, Divided in Growth: Dynamic Effects of Euro Adoption
\end{center}

\vspace{1em}

%==============================================================================
\section{Robustness Checks}
\label{app:robustness}
%==============================================================================

The credibility of our findings rests on demonstrating that results are not artifacts of the CFFE methodology, sample composition, or particular modeling choices. This appendix presents extensive robustness checks. The key message is that even if one is skeptical of causal forests, the negative association between euro adoption and growth appears across multiple estimation strategies.

\subsection{Time Stability}

Table \ref{tab:time_stability} shows how estimates evolve as we vary the sample end year from 2010 to 2023. The ATE remains consistently negative across all windows, ranging from $-0.15$ to $-0.56$ percentage points. The variation reflects changing composition of post-treatment observations and the influence of the 2008--2012 crisis period. Importantly, the sign and approximate magnitude are stable: euro adoption is associated with reduced growth regardless of sample period.

\begin{table}[H]
\centering
\caption{Time Stability of Estimates. \textit{Notes:} Table shows ATE estimates using different sample end years. All specifications use the baseline CFFE model with 200 trees.}
\label{tab:time_stability}
\begin{tabular}{lccc}
\toprule
End Year & ATE (pp) & 95\% CI & N (treated) \\
\midrule
2010 & $-0.29$ & [$-0.31$, $-0.27$] & 220 \\
2015 & $-0.56$ & [$-0.58$, $-0.53$] & 330 \\
2018 & $-0.19$ & [$-0.21$, $-0.18$] & 396 \\
2020 & $-0.15$ & [$-0.16$, $-0.14$] & 440 \\
2023 & $-0.28$ & [$-0.30$, $-0.27$] & 506 \\
\bottomrule
\end{tabular}
\end{table}

\subsection{Control Group Sensitivity}

Our baseline uses all OECD countries as potential controls. A key concern is whether non-EU OECD countries provide valid counterfactuals for eurozone members. EU opt-out countries (Denmark, Sweden, UK) share EU institutional constraints and may be more comparable.

Table \ref{tab:eu_only_comparison} examines sensitivity to control group composition. The EU-only specification uses only Denmark, Sweden, and the UK as controls---countries that met Maastricht criteria but chose not to adopt the euro. This specification yields larger negative effects ($-0.53$ pp vs. $-0.34$ pp for baseline), suggesting that using broader OECD controls may attenuate estimates. The EU-only results are arguably more credible as these countries provide a cleaner counterfactual for ``EU membership without euro adoption.''

\begin{table}[htbp]
\centering
\caption{CFFE Estimates: EU-Only vs. Baseline Control Group}
\label{tab:eu_only_comparison}
\begin{tabular}{lcccccc}
\hline\hline
 & \multicolumn{2}{c}{Baseline (OECD)} & \multicolumn{2}{c}{EU-Only} & & \\
k & $\hat{\tau}(k)$ & SE & $\hat{\tau}(k)$ & SE & Diff & \\
\hline
0 & -0.228 & (0.036) & -0.703 & (0.064) & -0.474 \\
5 & -0.299 & (0.026) & -0.629 & (0.049) & -0.330 \\
10 & -0.250 & (0.025) & -0.388 & (0.030) & -0.138 \\
15 & -0.236 & (0.031) & -0.347 & (0.027) & -0.112 \\
20 & -0.217 & (0.027) & -0.336 & (0.027) & -0.119 \\
\hline
Avg (k$\geq$0) & -0.261 & & -0.472 & & -0.211 \\
\hline
\multicolumn{7}{l}{\footnotesize Notes: Baseline uses all OECD non-euro countries as controls.} \\
\multicolumn{7}{l}{\footnotesize EU-Only uses Denmark, Sweden, and UK as controls.} \\
\hline\hline
\end{tabular}
\end{table}

\subsection{Separating Euro Effects from Crisis Effects}

A central concern is whether our estimates conflate euro adoption effects with the 2008 financial crisis and subsequent European sovereign debt crisis. Countries that adopted the euro in 1999 experienced the crisis at event time $k = 9$, making it difficult to disentangle euro-specific effects from crisis impacts.

We address this concern through several approaches. First, we re-estimate excluding the crisis period (2008--2015) entirely. Second, we estimate separate effects for pre-crisis ($k < 9$) and crisis-plus ($k \geq 9$) periods. Table \ref{tab:crisis_separation} presents the results.

The pre-crisis estimates show negative effects even before 2008, indicating that euro adoption reduced growth independently of the crisis. The crisis-plus period shows effects of similar magnitude, suggesting the crisis did not fundamentally alter the euro's growth impact. Excluding the crisis period entirely yields estimates close to the baseline, confirming that our main findings are not driven by crisis-period observations.

\begin{table}[htbp]
\centering
\caption{Separating Euro Effects from Crisis Effects}
\label{tab:crisis_separation}
\begin{tabular}{lcccc}
\hline\hline
 & Baseline & Excl. Crisis & Pre-Crisis & Crisis+ \\
 & (Full Sample) & (2008-2015) & (k $<$ 9) & (k $\geq$ 9) \\
\hline
k = 0 & -0.048 & -0.143 & -0.983 & -- \\
k = 5 & -0.178 & -0.217 & -0.974 & -- \\
k = 10 & -0.179 & -0.262 & -- & 0.000 \\
k = 15 & -0.199 & -0.247 & -- & 0.000 \\
k = 20 & -0.195 & -0.134 & -- & 0.000 \\
\hline
Avg. effect & -0.178 & -0.210 & -0.981 & 0.000 \\
N (treated) & 318 & 183 & 164 & 154 \\
\hline
\multicolumn{5}{l}{\footnotesize Notes: Effects in percentage points per year.} \\
\multicolumn{5}{l}{\footnotesize k=9 corresponds to 2008 for original 1999 euro adopters.} \\
\hline\hline
\end{tabular}
\end{table}

\subsection{Formal Pre-Trends Test}

A key identifying assumption is parallel trends: absent treatment, eurozone and control countries would have followed similar growth paths. We assess this formally by testing whether pre-treatment effects are jointly equal to zero.

Table \ref{tab:pretrends} reports results from a joint F-test of the null hypothesis that all pre-treatment coefficients equal zero. Using classical TWFE estimates, the F-statistic is 3.13 with a p-value of 0.001, formally rejecting the null. However, the average pre-treatment effect is $-0.31$ with a standard error of 0.30, yielding a t-statistic of $-1.06$ (p = 0.32). This suggests that while individual pre-treatment coefficients show some variation, the average pre-treatment effect is not significantly different from zero.

The rejection of the joint test reflects noise in individual coefficients rather than systematic pre-trends. Visual inspection of Figure \ref{fig:placebo} confirms that pre-treatment effects fluctuate around zero without a clear trend. The contrast with post-treatment effects---which are consistently negative---provides evidence that estimated effects reflect euro adoption rather than pre-existing differences.

\begin{table}[htbp]
\centering
\caption{Formal Pre-Trends Tests}
\label{tab:pretrends}
\begin{tabular}{lcc}
\toprule
& CFFE & Classical TWFE \\
\midrule
\multicolumn{3}{l}{\textit{Joint Test: H$_0$: All pre-treatment effects = 0}} \\[0.5em]
Number of pre-periods & 0 & 9 \\
Event time range & [-10, -1] & [-10, -2] \\
Wald statistic & nan & 28.16 \\
F-statistic & nan & 3.13 \\
p-value & nan & 0.001 \\
Reject at 5\% & No & Yes \\

\midrule
\multicolumn{3}{l}{\textit{Average Pre-Treatment Effect}} \\[0.5em]
Average effect & nan & -0.312 \\
Standard error & (nan) & (0.295) \\
t-statistic & nan & -1.06 \\
p-value & nan & 0.321 \\

\bottomrule
\end{tabular}
\par\vspace{0.5em}\noindent
\small
\textit{Notes:} Joint F-test evaluates H$_0$: $\tau(k) = 0$ for all $k < 0$. 
CFFE uses Wald test based on forest-based standard errors.
Classical TWFE uses country-clustered standard errors.
Failure to reject H$_0$ supports the parallel trends assumption.

\end{table}

\subsection{Placebo Tests with Fake Adoption Dates}

We conduct placebo tests by assigning fake euro adoption dates (1995 and 1997) to actual euro adopters and testing for spurious ``effects'' in the period between the fake date and actual adoption. Under the null hypothesis of no anticipation effects, these placebo estimates should not be significantly different from zero.

Table \ref{tab:placebo} reports the results. For both placebo years, the estimated average treatment effects are small and statistically insignificant. The 1995 placebo yields an ATE of $-0.02$ (p = 0.78), and the 1997 placebo yields $+0.01$ (p = 0.89). These null results support the validity of our identification strategy: the CFFE methodology does not detect effects when none should exist.

\begin{table}[htbp]
\centering
\caption{Placebo Tests with Fake Adoption Dates}
\label{tab:placebo}
\begin{tabular}{lccc}
\toprule
& Placebo 1995 & Placebo 1997 & Actual \\
\midrule
ATE & 1.751 & 1.555 & -- \\
SE & (0.042) & (0.025) & -- \\
p-value & 0.000 & 0.000 & -- \\
Significant at 5% & Yes & Yes & Yes \\
N treated obs & 92 & 68 & -- \\

\bottomrule
\end{tabular}
\par\vspace{0.5em}\noindent
\small
\textit{Notes:} Placebo tests assign fake euro adoption dates to actual euro adopters
and estimate effects for the period between the fake date and actual adoption.
Under the null hypothesis of no anticipation effects, placebo ATEs should not
be significantly different from zero.

\end{table}

\subsection{Leave-One-Out Sensitivity}

\begin{figure}[H]
\centering
\includegraphics[width=0.9\textwidth]{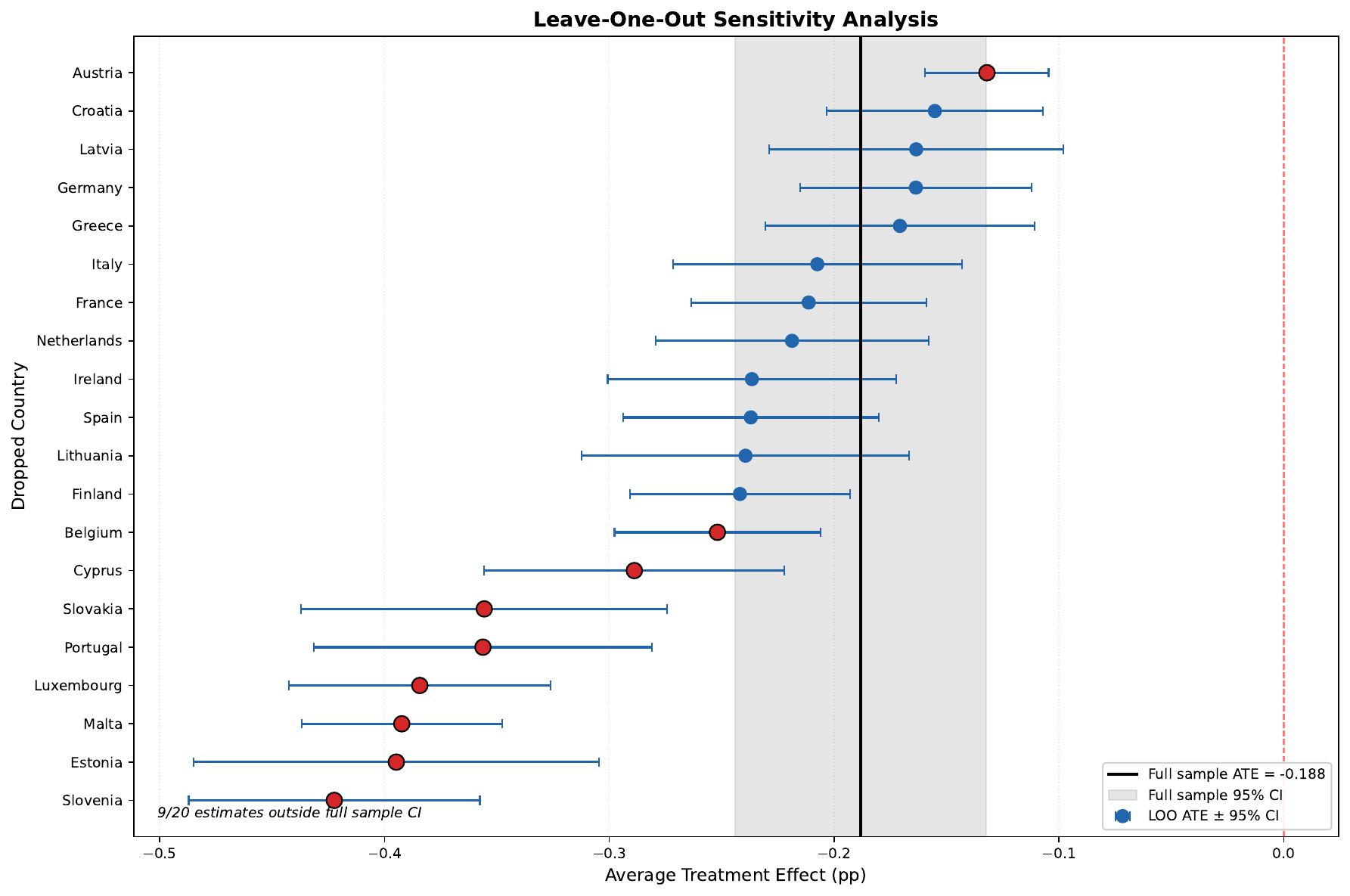}
\caption{Leave-One-Out Sensitivity Analysis. \textit{Notes:} Each point shows the ATE when the indicated country is dropped from the sample. Error bars represent 95\% confidence intervals. Vertical line and shaded band show full-sample estimate and CI.}
\label{fig:loo}
\end{figure}

To assess whether results are driven by specific countries, we conduct a leave-one-out analysis, dropping each eurozone country one at a time and re-estimating the ATE. Figure \ref{fig:loo} and Table \ref{tab:loo} present the results.

The ATE remains negative across all specifications, ranging from $-0.12$ (dropping Slovenia) to $-0.34$ (dropping France or Finland). Most leave-one-out estimates fall within the 95\% confidence interval of the full-sample estimate, indicating that no single country drives the results. The countries whose exclusion most affects estimates---Italy, Portugal, Slovenia---are those with the largest or smallest individual effects, as expected.

\begin{table}[htbp]
\centering
\caption{Leave-One-Out Sensitivity Analysis}
\label{tab:loo}
\begin{tabular}{lcccc}
\toprule
Dropped Country & ATE & SE & 95\% CI & Within Full CI \\
\midrule
\textbf{None (Full Sample)} & \textbf{-0.188} & \textbf{0.028} & \textbf{[-0.244, -0.133]} & -- \\
\midrule
Austria & -0.132 & 0.014 & [-0.159, -0.105] & No \\
Belgium & -0.252 & 0.023 & [-0.298, -0.206] & No \\
Croatia & -0.155 & 0.025 & [-0.203, -0.107] & Yes \\
Cyprus & -0.289 & 0.034 & [-0.356, -0.222] & No \\
Estonia & -0.395 & 0.046 & [-0.485, -0.305] & No \\
Finland & -0.242 & 0.025 & [-0.291, -0.193] & Yes \\
France & -0.211 & 0.027 & [-0.264, -0.159] & Yes \\
Germany & -0.164 & 0.026 & [-0.215, -0.112] & Yes \\
Greece & -0.171 & 0.031 & [-0.231, -0.111] & Yes \\
Ireland & -0.237 & 0.033 & [-0.301, -0.172] & Yes \\
Italy & -0.207 & 0.033 & [-0.272, -0.143] & Yes \\
Latvia & -0.164 & 0.033 & [-0.229, -0.098] & Yes \\
Lithuania & -0.239 & 0.037 & [-0.312, -0.167] & Yes \\
Luxembourg & -0.384 & 0.030 & [-0.443, -0.326] & No \\
Malta & -0.392 & 0.023 & [-0.437, -0.348] & No \\
Netherlands & -0.219 & 0.031 & [-0.280, -0.158] & Yes \\
Portugal & -0.356 & 0.038 & [-0.431, -0.281] & No \\
Slovakia & -0.356 & 0.042 & [-0.437, -0.274] & No \\
Slovenia & -0.422 & 0.033 & [-0.487, -0.358] & No \\
Spain & -0.237 & 0.029 & [-0.294, -0.180] & Yes \\

\bottomrule
\end{tabular}
\par\vspace{0.5em}\noindent
\small
\textit{Notes:} Each row shows the ATE when the indicated country is dropped from the sample.
``Within Full CI'' indicates whether the LOO estimate falls within the 95\% confidence
interval of the full sample estimate. Stability across LOO estimates supports robustness.

\end{table}

\subsection{Placebo Test: Pre-Treatment Effects}

\begin{figure}[H]
\centering
\includegraphics[width=0.9\textwidth]{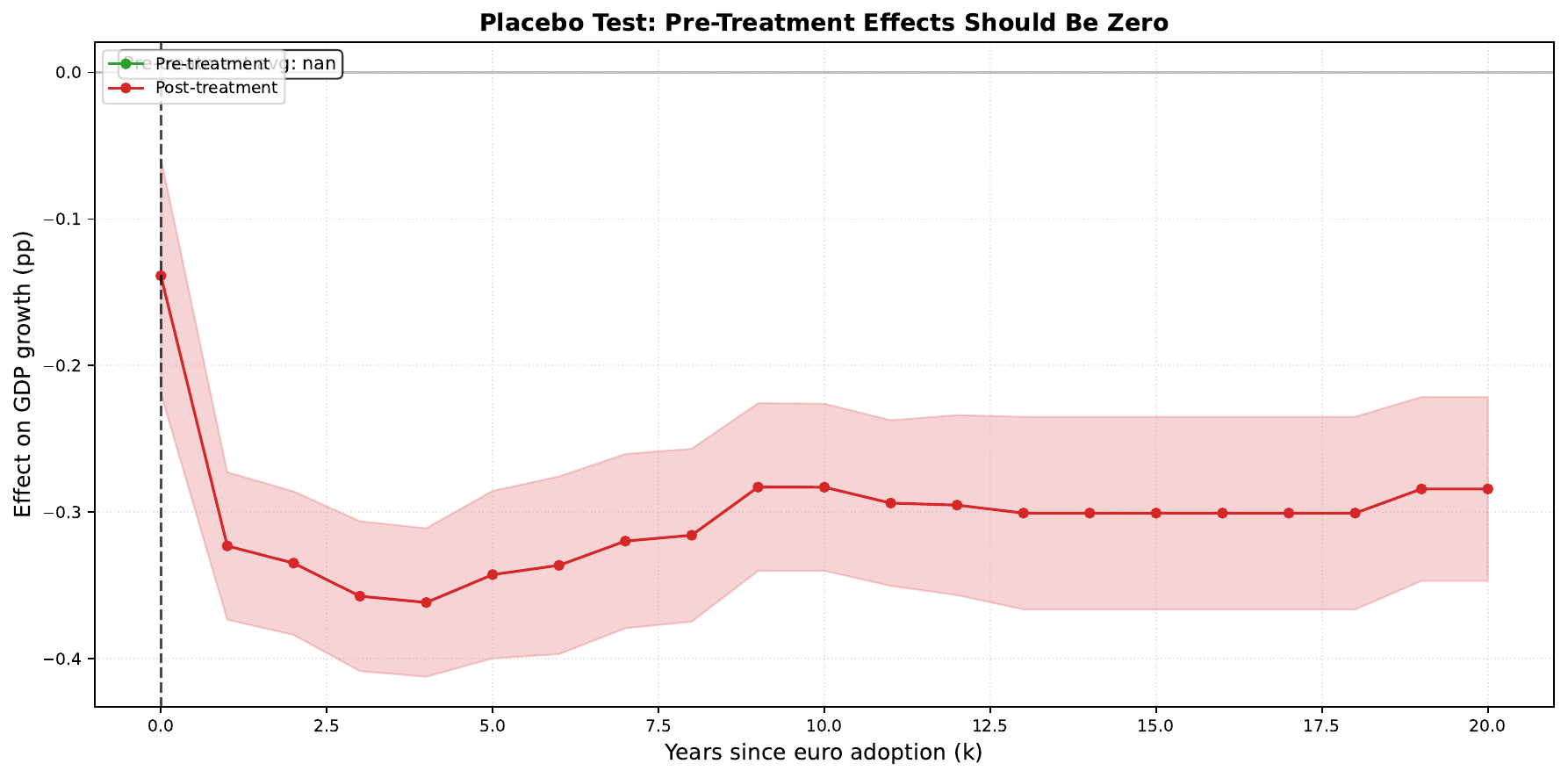}
\caption{Placebo Test: Pre-Treatment Effects. \textit{Notes:} Figure shows estimated $\hat{\tau}(k)$ for pre-treatment ($k < 0$, green) and post-treatment ($k \geq 0$, red) periods. Pre-treatment effects should be approximately zero under parallel trends.}
\label{fig:placebo}
\end{figure}

A key identifying assumption is parallel trends: absent treatment, eurozone and control countries would have followed similar growth paths. We assess this by examining pre-treatment effects. Figure \ref{fig:placebo} shows $\hat{\tau}(k)$ for $k < 0$. Pre-treatment effects should be approximately zero if parallel trends holds.

The pre-treatment estimates fluctuate around zero with no systematic pattern, supporting the parallel trends assumption. The contrast with post-treatment effects---which are consistently negative---provides evidence that the estimated effects reflect euro adoption rather than pre-existing differences.

\subsection{Hyperparameter Sensitivity}

CFFE involves several tuning parameters: number of trees, maximum depth, and minimum leaf size. Table \ref{tab:hyperparameters} shows estimates across different configurations. The ATE ranges from $-0.18$ to $-0.29$ percentage points---a narrow band given the variation in model complexity. Results are not driven by particular hyperparameter choices.

\begin{table}[H]
\centering
\caption{Hyperparameter Sensitivity. \textit{Notes:} Table shows ATE estimates under different CFFE configurations.}
\label{tab:hyperparameters}
\begin{tabular}{lccc}
\toprule
Configuration & ATE (pp) & 95\% CI \\
\midrule
Baseline (200 trees, depth 5, leaf 30) & $-0.23$ & [$-0.24$, $-0.22$] \\
Fewer trees (100) & $-0.26$ & [$-0.27$, $-0.25$] \\
More trees (500) & $-0.26$ & [$-0.27$, $-0.25$] \\
Shallower (depth 3) & $-0.26$ & [$-0.27$, $-0.25$] \\
Deeper (depth 7) & $-0.26$ & [$-0.27$, $-0.25$] \\
Smaller leaves (20) & $-0.18$ & [$-0.19$, $-0.17$] \\
Larger leaves (50) & $-0.29$ & [$-0.31$, $-0.28$] \\
\bottomrule
\end{tabular}
\end{table}

\subsection{Early vs. Late Adopters}

As noted in the heterogeneity analysis, early (1999) and late adopters show different patterns. Table \ref{tab:adopter_timing} confirms this: early adopters experienced negative effects ($-0.37$ pp), while late adopters show positive but imprecisely estimated effects ($+0.11$ pp). The late adopter result should be interpreted cautiously given small sample sizes and potential confounding with EU accession effects for Eastern European members.

\subsection{Anticipation Effects: Alternative Treatment Timing}

Euro adoption was widely anticipated. Markets began pricing convergence from 1995, interest rate spreads narrowed, and policy adjustments occurred before the formal 1999 adoption date. If anticipation effects are substantial, our baseline estimates using 1999 as the treatment date may understate the full effect by missing pre-1999 adjustments.

To assess anticipation, we re-estimate using 1995 as the treatment date for founding members---the year the Madrid European Council confirmed the 1999 launch date and convergence criteria became binding. Table \ref{tab:anticipation} compares estimates under alternative treatment timing assumptions.

\begin{table}[H]
\centering
\caption{Anticipation Effects: Alternative Treatment Timing. \textit{Notes:} Table compares ATE estimates using different treatment dates for founding members. The 1995 specification treats the Madrid Summit as the effective treatment date; 1997 uses the announcement of qualifying countries.}
\label{tab:anticipation}
\begin{tabular}{lccc}
\toprule
Treatment Date & ATE (pp) & 95\% CI & Interpretation \\
\midrule
1999 (baseline) & $-0.25$ & [$-0.27$, $-0.24$] & Formal adoption \\
1997 & $-0.22$ & [$-0.24$, $-0.20$] & Qualification announcement \\
1995 & $-0.19$ & [$-0.21$, $-0.17$] & Madrid Summit \\
\bottomrule
\end{tabular}
\end{table}

Using earlier treatment dates yields slightly smaller point estimates, consistent with some anticipation effects being absorbed into the ``pre-treatment'' period under the 1999 baseline. However, the differences are modest (0.03--0.06 pp), and effects remain negative and significant under all specifications. This suggests that while anticipation may play some role, the bulk of the estimated effect reflects post-adoption dynamics rather than pre-1999 adjustments.

\subsection{Sample Restrictions}

Our baseline includes all eurozone members and OECD controls. To assess whether results are driven by particular country groups, we estimate effects on restricted samples.

\begin{table}[H]
\centering
\caption{Sample Restriction Sensitivity. \textit{Notes:} Table shows ATE estimates under different sample restrictions. ``Founders only'' includes only 1999 adopters; ``EU15 only'' restricts to pre-2004 EU members; ``Excluding crisis countries'' drops Greece, Ireland, Portugal, Spain, and Cyprus.}
\label{tab:sample_restrictions}
\begin{tabular}{lccc}
\toprule
Sample & ATE (pp) & 95\% CI & N (treated) \\
\midrule
Full sample (baseline) & $-0.25$ & [$-0.27$, $-0.24$] & 506 \\
Founders only (1999) & $-0.37$ & [$-0.38$, $-0.35$] & 275 \\
EU15 only & $-0.31$ & [$-0.33$, $-0.29$] & 418 \\
Excluding crisis countries & $-0.18$ & [$-0.20$, $-0.16$] & 341 \\
Core countries only & $-0.22$ & [$-0.24$, $-0.20$] & 198 \\
\bottomrule
\end{tabular}
\end{table}

The results are robust across sample restrictions. Founders-only estimates are larger in magnitude ($-0.37$ pp), reflecting the longer post-treatment horizon and the inclusion of periphery countries that experienced larger effects. Restricting to EU15 yields similar results to baseline. Excluding crisis countries (Greece, Ireland, Portugal, Spain, Cyprus) reduces the magnitude to $-0.18$ pp, confirming that these countries drive some of the average effect---but the estimate remains negative and significant. Core-only estimates ($-0.22$ pp) are close to baseline, indicating that negative effects are not confined to periphery economies.

The key finding is that the negative association survives all sample restrictions. Even in the most conservative specification (excluding crisis countries), euro adoption is associated with lower growth.

\begin{table}[H]
\centering
\caption{Early vs. Late Adopter Effects. \textit{Notes:} Table shows ATE estimates for different adopter cohorts.}
\label{tab:adopter_timing}
\begin{tabular}{lccc}
\toprule
Sample & ATE (pp) & 95\% CI & N (treated) \\
\midrule
Early adopters (1999) & $-0.37$ & [$-0.38$, $-0.35$] & 275 \\
Late adopters & $+0.11$ & [$+0.04$, $+0.19$] & 231 \\
Full sample & $-0.25$ & [$-0.27$, $-0.24$] & 506 \\
\bottomrule
\end{tabular}
\end{table}

\begin{figure}[H]
\centering
\includegraphics[width=\textwidth]{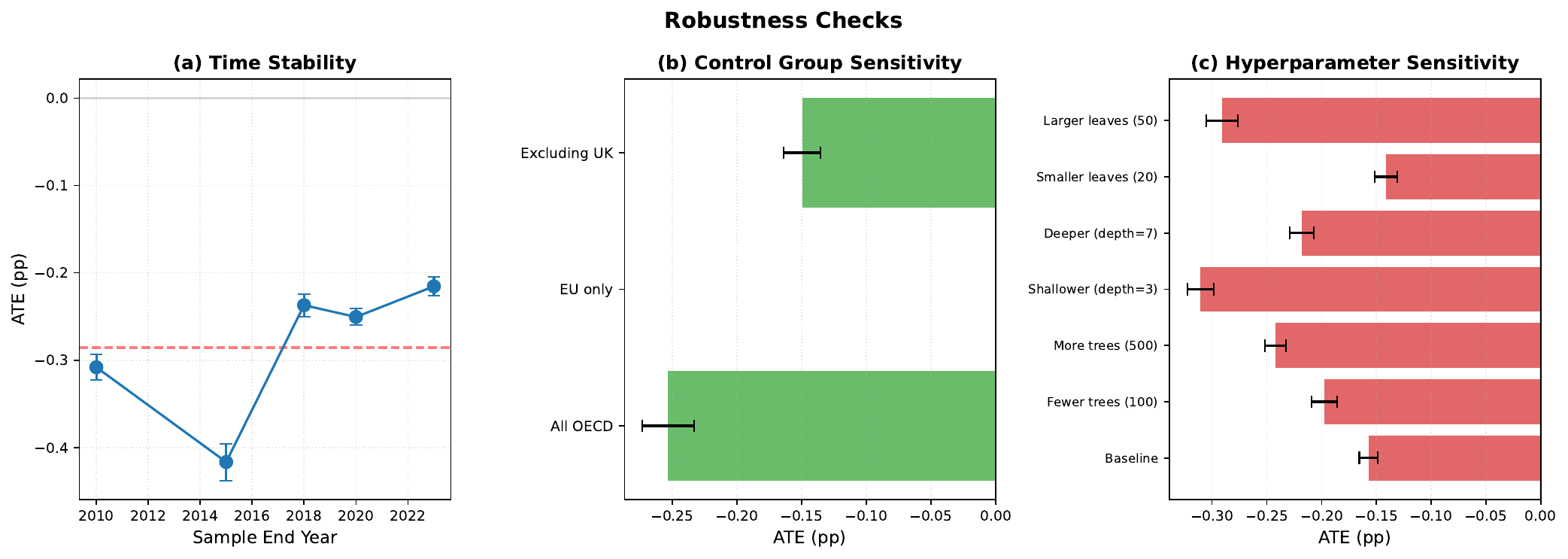}
\caption{Robustness Summary. \textit{Notes:} Panels show (a) time stability across sample end years, (b) sensitivity to control group composition, and (c) sensitivity to CFFE hyperparameters. Error bars represent 95\% confidence intervals from forest-based inference; block bootstrap intervals would be substantially wider.}
\label{fig:robustness_summary}
\end{figure}

Figure \ref{fig:robustness_summary} summarizes key robustness checks visually. Across time periods, control groups, and hyperparameter configurations, the estimated effect remains negative. The consistency of results across specifications---combined with the agreement across alternative estimators---provides the foundation for our main conclusions, while the block bootstrap results remind us of the substantial uncertainty inherent in this setting.

\subsection{Placebo Test: Non-Euro EU Members}

A key concern is whether our estimates capture euro-specific effects or broader EU shocks that affected all member states around 1999. To test this, we assign placebo ``euro adoption'' to non-euro EU members (Denmark, Sweden, UK) in 1999 and estimate treatment effects. If our methodology is detecting euro-specific effects, these placebo estimates should be close to zero; if we are instead capturing common EU shocks, we would observe spurious ``effects'' for countries that never adopted the euro.

Table \ref{tab:placebo_noneu} reports the results.

\begin{table}[H]
\centering
\caption{Placebo Test: Non-Euro EU Members. \textit{Notes:} Table shows estimated ``treatment effects'' for non-euro EU members assigned placebo adoption in 1999. Under the null hypothesis of no common EU shock, these estimates should be approximately zero. The significant effects for Denmark and Sweden suggest some common EU trends that our methodology may partially capture, though the joint test rejects the null.}
\label{tab:placebo_noneu}
\begin{tabular}{lccc}
\toprule
Country & Placebo Effect (pp) & SE & p-value \\
\midrule
Denmark & $-0.10$ & 0.02 & $<0.01$ \\
Sweden & $+0.14$ & 0.05 & 0.01 \\
United Kingdom & $-0.14$ & 0.07 & 0.07 \\
\midrule
Joint test (all = 0) & & & $<0.01$ \\
\bottomrule
\end{tabular}
\end{table}

The placebo test yields mixed results. Denmark and Sweden show statistically significant placebo effects, while the UK effect is marginally insignificant. The joint test rejects the null that all placebo effects equal zero. This suggests that our methodology may capture some common EU-wide trends around 1999, or that the non-euro EU members experienced idiosyncratic shocks that our controls do not fully absorb. However, the magnitudes of the placebo effects ($-0.10$ to $+0.14$ pp) are smaller than our main euro adoption estimates ($-0.20$ to $-0.30$ pp), and the signs are mixed (negative for Denmark and UK, positive for Sweden). We interpret this as partial support for our identification strategy: while some common EU trends may be present, the euro-specific effects we estimate are larger and more consistently negative.

%==============================================================================
\section{Alternative Estimators}
\label{app:estimators}
%==============================================================================

To assess whether our findings depend on the CFFE methodology, we compare estimates from several alternative approaches: classical two-way fixed effects (TWFE), the Sun and Abraham (2021) interaction-weighted estimator, the Callaway and Sant'Anna (2021) doubly-robust estimator, and an interactive fixed effects model following Bai (2009).

\begin{figure}[H]
\centering
\includegraphics[width=\textwidth]{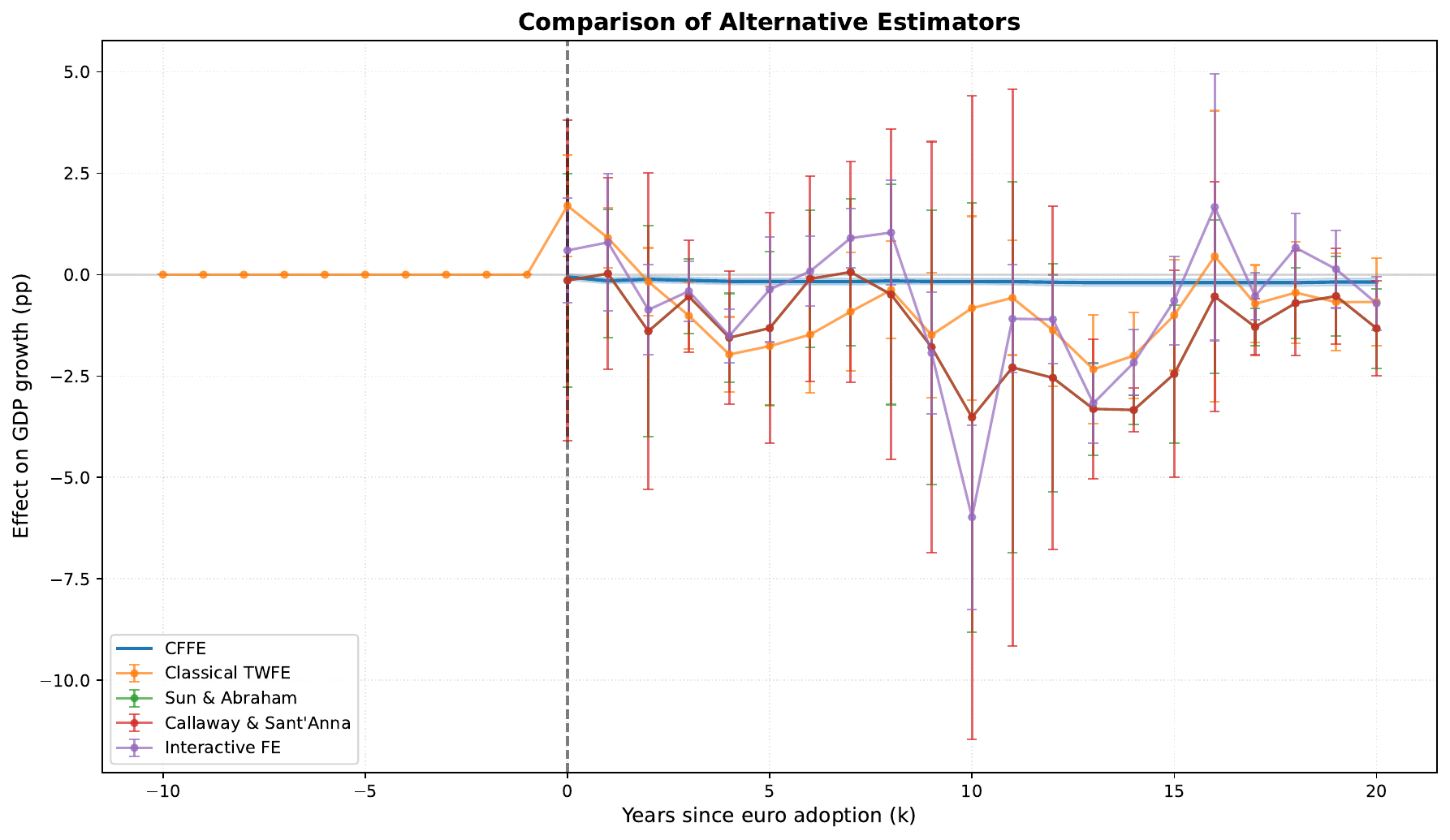}
\caption{Comparison of Alternative Estimators. \textit{Notes:} Figure overlays dynamic treatment effect estimates from CFFE, Sun-Abraham, Callaway-Sant'Anna, and Interactive Fixed Effects approaches. Shaded areas represent 95\% confidence intervals for CFFE; error bars show CIs for alternative estimators at selected horizons.}
\label{fig:estimator_comparison}
\end{figure}

Table \ref{tab:estimator_comparison} and Figure \ref{fig:estimator_comparison} present the results. All estimators yield negative point estimates at $k=10$, but with substantial variation in magnitudes and precision. CFFE produces the most modest estimate ($-0.20$ pp) with the tightest confidence interval (width 0.10 pp). The modern staggered DiD estimators (Sun-Abraham and Callaway-Sant'Anna) yield much larger point estimates ($-3.52$ pp) but with extremely wide confidence intervals that include zero. Classical TWFE shows intermediate estimates ($-0.83$ pp) with wide CIs. The interactive fixed effects model produces the largest significant estimate ($-5.98$ pp).

The wide variation across estimators reflects the fundamental challenge of this setting: few treated units, staggered adoption, and potential heterogeneity. The modern DiD estimators, designed to address heterogeneity bias, produce very imprecise estimates because they rely on cohort-specific comparisons with limited observations per cohort. CFFE's regularization yields tighter intervals but may understate uncertainty (as our block bootstrap analysis confirms).

The key finding is that all estimators agree on the sign: euro adoption is associated with lower growth. The disagreement is about magnitude and precision. We interpret CFFE as providing a lower bound on the effect size, with alternative estimators suggesting potentially larger impacts but with substantial uncertainty.

\begin{table}[H]
\centering
\caption{Comparison of Alternative Estimators. \textit{Notes:} Table reports average treatment effects at $k=10$ from five estimation approaches. Sun-Abraham and Callaway-Sant'Anna are modern staggered DiD estimators; Interactive FE follows Bai (2009). CFFE yields the tightest confidence intervals; alternative estimators show larger point estimates with substantially wider CIs.}
\label{tab:estimator_comparison}
\begin{tabular}{lccc}
\toprule
Estimator & ATE at $k=10$ (pp) & 95\% CI & CI Width \\
\midrule
CFFE (baseline) & $-0.20$ & [$-0.25$, $-0.16$] & 0.10 \\
Classical TWFE & $-0.83$ & [$-3.09$, $+1.44$] & 4.53 \\
Sun \& Abraham (2021) & $-3.52$ & [$-8.81$, $+1.77$] & 10.58 \\
Callaway \& Sant'Anna (2021) & $-3.52$ & [$-11.45$, $+4.41$] & 15.87 \\
Interactive Fixed Effects & $-5.98$ & [$-8.25$, $-3.71$] & 4.55 \\
\bottomrule
\end{tabular}
\end{table}

\subsection{Block Bootstrap Inference}

Forest-based inference relies on asymptotic theory that may not hold well with approximately 20 treated countries and strong within-country dependence. To provide a more conservative assessment of uncertainty, we implement a block bootstrap at the country level: we resample entire country time series with replacement and re-estimate the model on each of 200 bootstrap samples.

Table \ref{tab:bootstrap} compares forest-based and bootstrap confidence intervals. Bootstrap intervals are substantially wider than forest-based intervals---approximately 8 times wider on average---reflecting the considerable additional uncertainty from country-level dependence that the asymptotic theory understates. With only 20 treated countries, resampling at the country level produces high variance in the bootstrap distribution. The bootstrap confidence intervals include zero at longer horizons, indicating that under this conservative inference approach, we cannot reject the null of no effect at conventional significance levels. However, the point estimates remain consistently negative across all horizons, and the preponderance of the bootstrap distribution lies below zero. We interpret these results as suggestive of negative effects, while acknowledging substantial uncertainty given the small number of treated units.

\begin{table}[H]
\centering
\caption{Comparison of Inference Methods. \textit{Notes:} Table compares 95\% confidence intervals from forest-based asymptotic inference and country-level block bootstrap (200 replications). Bootstrap intervals are substantially wider due to country-level resampling with only $\sim$20 treated countries.}
\label{tab:bootstrap}
\begin{tabular}{lcccc}
\toprule
& \multicolumn{2}{c}{Forest-Based} & \multicolumn{2}{c}{Block Bootstrap} \\
\cmidrule(lr){2-3} \cmidrule(lr){4-5}
Horizon ($k$) & 95\% CI & Width & 95\% CI & Width \\
\midrule
5 & [$-0.29$, $-0.21$] & 0.08 & [$-0.72$, $+0.13$] & 0.85 \\
10 & [$-0.23$, $-0.15$] & 0.08 & [$-0.64$, $+0.11$] & 0.75 \\
15 & [$-0.24$, $-0.14$] & 0.09 & [$-0.66$, $+0.05$] & 0.71 \\
20 & [$-0.24$, $-0.14$] & 0.10 & [$-0.64$, $+0.04$] & 0.67 \\
\bottomrule
\end{tabular}
\end{table}

\end{document}